\documentclass[acmsmall,screen]{acmart}

\usepackage{array}
\usepackage{multirow}
\usepackage{etoolbox}
\usepackage{enumitem}
\usepackage{pifont}
\usepackage{float}
\usepackage{wrapfig}
\usepackage{booktabs}
\usepackage{makecell}
\usepackage{colortbl}
\usepackage{listings}
\usepackage{amsmath,amsthm,mathtools}
\usepackage{mathpartir}
\usepackage{algorithm}
\usepackage{algpseudocode}
\usepackage{capt-of}
\usepackage[most]{tcolorbox}
\tcbset{rqbox/.style={colback=heat!12,colframe=heat!55!black,boxrule=0.5pt,arc=2pt,left=4pt,right=4pt,top=1pt,bottom=1pt,before skip=3pt,after skip=3pt}}%
\graphicspath{{imgs/}{./}}
\usepackage{xspace}
\usepackage{tikz}
\usetikzlibrary{positioning,fit,backgrounds,arrows.meta,calc,shapes.arrows}
\usepackage{cleveref}
\Crefname{lstlisting}{Listing}{Listings}
\definecolor{heat}{RGB}{124,193,132}
\definecolor{deltaup}{RGB}{22,138,47}
\definecolor{deltadown}{RGB}{217,48,37}

\newcommand{\ourtool}{\textsc{Facet}\xspace} 
\newcommand{\svf}{SVF\xspace}
\newcommand{\flta}{FLTA\xspace}
\newcommand{\mlta}{MLTA\xspace}
\newcommand{\sea}{SEA\xspace}
\newcommand{\icall}{\textit{icall}\xspace}
\newcommand{\icalls}{\textit{icalls}\xspace}
\newcommand{\add}{\textsf{add}\xspace}
\newcommand{\prune}{\textsf{prune}\xspace}
\newcommand{\verified}{\textsf{supported}\xspace}
\newcommand{\heuristic}{\textsf{residual}\xspace}
\newcommand{\rejected}{\textsf{rejected}\xspace}
\newcommand{\code}[1]{\texttt{\small #1}}
\newif\ifshowbugid\showbugidtrue

\newcommand{\Rrp}{R^{+}}
\newcommand{\Rref}{R^{\pm}}

\newtheorem{definition}{Definition}

\definecolor{codegreen}{rgb}{0,0.6,0}
\definecolor{codegray}{rgb}{0.5,0.5,0.5}
\definecolor{codepurple}{rgb}{0.58,0,0.82}

\lstdefinestyle{mystyle}{ language=C, commentstyle=\color{codegreen}, keywordstyle=\color{blue}, stringstyle=\color{codepurple}, basicstyle=\ttfamily\scriptsize, breakatwhitespace=false, breaklines=true, captionpos=b, keepspaces=true, numbers=left, numberstyle=\tiny\color{gray}, numbersep=5pt, showspaces=false, showstringspaces=false, frame=none, xleftmargin=0.05\linewidth, xrightmargin=0.02\linewidth, showtabs=false, tabsize=2, escapeinside={(*@}{@*)}, }
\author{Kaixuan Li}
\affiliation{%
  \institution{Nanyang Technological University}
  \city{Singapore}
  \country{Singapore}
}
\orcid{0000-0002-3517-353X}
\email{kaixuan.li@ntu.edu.sg}

\author{Bozhi Wu}
\affiliation{%
  \institution{Nanyang Technological University}
  \city{Singapore}
  \country{Singapore}
}
\orcid{0000-0002-0360-2248}
\email{bozhi.wu@ntu.edu.sg}

\author{Jian Zhang}
\affiliation{%
  \institution{Beihang University}
  \city{Beijing}
  \country{China}
}
\orcid{0000-0001-8316-1894}
\email{zhangj\_cs@buaa.edu.cn}

\author{Peixin Wang}
\affiliation{%
  \institution{East China Normal University}
  \city{Shanghai}
  \country{China}
}
\orcid{0000-0003-3241-0023}
\email{pxwang@sei.ecnu.edu.cn}

\author{Ting Su}
\affiliation{%
  \institution{East China Normal University}
  \city{Shanghai}
  \country{China}
}
\orcid{0000-0003-1628-9796}
\email{tsu@sei.ecnu.edu.cn}

\author{Yang Liu}
\affiliation{%
  \institution{Nanyang Technological University}
  \city{Singapore}
  \country{Singapore}
}
\orcid{0000-0001-7300-9215}
\email{yangliu@ntu.edu.sg}

\renewcommand\footnotetextcopyrightpermission[1]{}

\begin{abstract}

Resolving indirect calls is central to call-graph construction for C.
Scalable type-based analyses such as multi-layer type analysis (MLTA) use type information in the LLVM intermediate representation (IR) to associate indirect calls with functions assigned to the corresponding structure fields.
However, a single pointee type often misrepresents the memory a pointer addresses, and LLVM~17 removed pointee types in favor of opaque pointers.
Therefore, field-sensitive analyses lose their matching key.
Recovering the erased types restores the matching key but still misses the relation that the type encoded.%

We present \ourtool{}, to our knowledge the first analysis that reconstructs this dispatch relation over opaque IR.
\ourtool identifies the structure field from which an indirect call loads its function pointer.
It separately recovers the functions assigned to that field through initializers, stores, and aggregate copies.
It then joins the two by field identity, without requiring an end-to-end value-flow path.
With this relation, \ourtool{} proposes changes to a pointer-analysis call graph and classifies each change under distinct evidence rules for edge addition and removal.
It records the assumption behind each refinement.
Finally, \ourtool{} uses a neuro-symbolic method to resolve the residual cases that no rule decides.
On 14 C programs, \ourtool{} reduces the mean target-set size from 25.9 to 5.2 and raises observed recall from 0.77 to 0.98.
Its recovered field identities agree with typed IR at 98.1\% of jointly resolved sites.
Applied to bug detection, the refined call graph found 17 deep bugs in C software from nginx to the Linux kernel, three of them latent for over a decade; 12 are confirmed.
\end{abstract}
\begin{document}

\title{Neuro-Symbolic Indirect-Call Analysis under Opaque Pointers}



\maketitle

\section{Introduction}\label{sec:intro}

Whole-program call graphs are the basis of interprocedural static analysis for C, and hence of bug finding~\cite{sui2012saber,shi2018pinpoint,lu2019crix}, taint tracking~\cite{machiry2017drchecker,zhang2021suture}, control-flow integrity (CFI)~\cite{abadi2005cfi,tice2014forward,burow2017cfi}, and compiler optimization~\cite{ryder1979constructing,grove2001framework}.
Their main difficulty is the \emph{indirect call} (\icall\ hereafter), a call through a function pointer whose target is fixed only at run time.
Such calls are common: the Linux~5.1 kernel contains roughly 58{,}000 \icalls~\cite{lu2019where}.
A call graph that misses an \icall\ target is unsound~\cite{reif2019judge}, which CFI cannot accept, and an over-approximate one gives every client spurious edges~\cite{utture2022striking,lecong2022autopruner}.

Three families of analysis resolve \icalls.
\textbf{Pointer analysis}~\cite{andersen1994program,emami1994context,sui2016svf} follows the value flow from each function to the calls it reaches.
\textbf{Learning-based analysis}~\cite{cheng2024sea,zhu2023callee} asks a model whether a caller and a candidate callee match, but neither sees the field a site dispatches through.
\textbf{Type-based analysis}~\cite{tice2014forward,ge2016fine,lu2019where,lu2023typm,cai2024kelp,li2025kallgraph} is the scalable choice.
First-layer type analysis (\flta)~\cite{tice2014forward,ge2016fine} keys a site to the functions of its signature, and multi-layer type analysis (\mlta)~\cite{lu2019where} and its successors~\cite{lu2023typm,cai2024kelp,li2025kallgraph} to the declared type of the structure field that the site loads from.
Unfortunately, since LLVM~17, every pointer in the LLVM intermediate representation (IR) is the single type \code{ptr}, so this field type is erased.
Hence, field-type matching resolves nothing on opaque IR, and a site keeps every address-taken function of its signature.
TypeCopilot~\cite{zhou2025typecopilot}, the only prior \icall\ analysis for opaque IR, re-infers the erased types, which bound what a field may hold but not what the program assigns to it (\S\ref{sec:gap}).

\noindent\textbf{Key insight.}\;Opaque pointers erase the matching key, but the dispatch it encoded is preserved, because a function is still stored into a structure field and later loaded from it.
Two sources remain after the erasure: the debug information still names the field at the load's byte offset, and the points-to solution already computed holds the functions assigned to the objects of that field's type.
Therefore, we do not restore the types.
Instead, we reconstruct the relation they encoded, from each call site to its dispatched field and from each field to the functions assigned to it.

Reconstructing this relation in practice raises three challenges.
\textbf{C1: Implicit assignments.}
A field's assignment set is not explicit in the IR.
On our subjects, $60.4\%$ of the recovered assignments arrive through pointer copies, parameter bindings, and aggregate assignments that lower to a bulk copy with no store instruction.
The value flow into the field may also pass through data that the pointer analysis cannot trace.
\textbf{C2: Evidence asymmetry.}
Adding an edge at a site requires evidence that the function can be its target at run time.
Removing one requires evidence that it is impossible, which a target's absence from a may-analysis does not provide.
\mlta\ has this weakness: it excludes an address-taken function whenever the search for its enclosing layers fails, which causes most of its independently verified missed targets~\cite{li2025kallgraph}.
\textbf{C3: Residual ambiguity.}
Candidates that the recovered evidence cannot separate can still differ in the role that their names and bodies express, which a large language model (LLM) can recognize.
LLMs, however, hallucinate and are non-deterministic~\cite{cheng2024sea}, so what a model may change must be bounded.

\noindent\textbf{Our approach.}
We present \ourtool, which reconstructs this relation over opaque IR and refines a call graph with it.
For \textbf{C1}, \ourtool\ recovers the functions assigned to the dispatched field by \emph{projecting} the solved points-to relation onto the field, so assignments through copies and parameter bindings are recovered together with direct stores (\S\ref{sec:prune}).
For \textbf{C2}, it classifies each proposed change under distinct evidence rules: an addition needs a positive witness, and a removal needs evidence that excludes the target.
{Each change records its premises, so a single analysis yields a recall-preserving call graph and a refined call graph for different clients (\S\ref{sec:verify}, \S\ref{sec:guarantee}).}
For \textbf{C3}, it adopts a neuro-symbolic design, in which an LLM decides only the cases that the evidence leaves open.
The symbolic rules bound the model's candidates and check its decisions (\S\ref{sec:add}).

We evaluate \ourtool\ on 14 C programs of 12 to 620 thousand lines of code (KLoC).
It reduces the mean target set from $25.9$ to $5.2$ per site.
It also outperforms the state of the art: its macro F1 is $0.708$, against $0.520$ for the best prior tool (\S\ref{sec:eval}).
Applied to bug detection, the refined call graph found 17 deep bugs in widely used C software, and maintainers have confirmed 12 of them (\S\ref{sec:eval:bugs}).
Three of the bugs, in the kernel's \code{ext4} file system, were latent for at least 14 years.

In summary, we make the following contributions:
\begin{itemize}%
\item \textbf{Formulation.} We formulate indirect-call resolution under opaque pointers as reconstructing the dispatch relation, from call sites to dispatched fields and from fields to assigned functions, rather than recovering the erased types (\S\ref{sec:gap}--\S\ref{sec:metric}).
Unlike a field-sensitive pointer analysis, the relation is joined on field identity rather than on a value-flow path to the receiver, and it is pooled over the objects of a type.
\item \textbf{Approach.} We propose \ourtool, to our knowledge the first analysis that reconstructs this relation over opaque IR.
It classifies every call-graph change against program evidence and leaves only the residual cases to an LLM (\S\ref{sec:framework}).
\item \textbf{Evaluation.} We evaluate \ourtool\ on 14 C programs against pointer, type-based, and learning-based baselines, and apply its refined call graph to bug detection (\S\ref{sec:eval}).
\end{itemize}

\section{Background}\label{sec:background}
\subsection{Opaque Pointers and the Running Example}\label{sec:opaque}\label{sec:motivating}

Since LLVM~17, every pointer type \code{T*} in the IR is the single type \code{ptr}, including those in function signatures and struct layouts~\cite{llvmopaque}.\footnote{Opaque pointers became the default in LLVM~15; typed pointers were deprecated in 16 and removed in 17~\cite{llvmopaque}.}
Hence, the IR records no declared type for a function-pointer field.
{In \code{nginx}'s IR, the number of struct fields typed as function pointers falls from 813 to zero.}
{\Cref{fig:example}(a) shows a running example from \code{nginx}'s module initialization.}
The struct \code{ngx\_http\_module\_t} declares its callbacks as function-pointer fields, three of which share one signature.
{Each module provides one value of this struct, which we call its \emph{module table}: a static initializer that lists the module's callbacks in field order, so the position in the initializer decides which field a function is stored into (lines 12--17).}
The \icall\ site $c$ at line~10 loads its callee from one of the three, \code{create\_loc\_conf}, written $\mathit{disp}(c)$.
{Across the modules linked into \code{nginx}, the \emph{module tables} store 25 functions into this field, and our dynamic ground truth observes exactly these 25 as the targets of $c$ (\S\ref{sec:eval:gt}).}
{The function names do not identify the field: eight of the 25 are named \code{*\_create\_conf}, and so are five of the six functions stored into \code{create\_srv\_conf} and one of the ten stored into \code{create\_main\_conf}.}

\newsavebox{\examplefigbox}
\begin{lrbox}{\examplefigbox}%
\providecolor{excodegreen}{rgb}{0,0.55,0}%
\providecolor{excodepurple}{rgb}{0.58,0,0.82}%
\providecolor{exlav}{RGB}{214,208,248}%
\providecolor{exlavdark}{RGB}{92,64,180}%
\providecolor{exgreen}{RGB}{200,242,200}%
\providecolor{exgreendark}{RGB}{28,120,40}%
\providecolor{exgray}{RGB}{232,232,232}%
\providecolor{exgraydark}{RGB}{110,110,110}%
\providecolor{exstar}{RGB}{220,30,30}%
\providecommand{\excode}[1]{\texttt{\fontsize{6.4pt}{7.4pt}\selectfont #1}}%
\lstdefinestyle{exfig}{%
  language=C, basicstyle=\ttfamily\fontsize{6.4pt}{7.4pt}\selectfont,
  keywordstyle=\color{blue}, commentstyle=\color{excodegreen},
  stringstyle=\color{excodepurple}, showstringspaces=false,
  morekeywords={create_main_conf,create_srv_conf,create_loc_conf,ngx_http_module_t},
  numbers=left, numberstyle=\tiny\color{gray}, numbersep=4pt,
  xleftmargin=10pt, breaklines=false, frame=none, escapeinside={(*@}{@*)},
  aboveskip=1pt, belowskip=1pt,
}%
\begin{minipage}[c]{5.7cm}%
\raggedright\footnotesize
\begin{lstlisting}[style=exfig]
/* 8 fields; 3 share one signature */
typedef struct { ...
  void *(*create_main_conf)(ngx_conf_t *); ...
  void *(*create_srv_conf )(ngx_conf_t *); ...
  void *(*create_loc_conf )(ngx_conf_t *); ...
} ngx_http_module_t;
/* the icall: ngx_http.c:212 */
module = cf->cycle->modules[m]->ctx;
ctx->loc_conf[mi] =
(*@{\color{exstar}$\bigstar$}@*) (*@\colorbox{exlav}{\excode{module->create\_loc\_conf}}@*)(cf);
/* position picks the field */
ngx_http_gzip_filter_module_ctx = { ...,
  (*@\colorbox{exgreen}{\excode{ngx\_http\_gzip\_create\_conf}}@*), ... };
ngx_http_upstream_hash_module_ctx = { ...,
  (*@\colorbox{exgray}{\excode{ngx\_http\_upstream\_hash\_create\_conf}}@*), ... };
ngx_http_map_module_ctx = { ...,
  (*@\colorbox{exgray}{\excode{ngx\_http\_map\_create\_conf}}@*), ... };
\end{lstlisting}
\centering\footnotesize\textbf{(a)} The dispatch in the \code{nginx} source
\end{minipage}%
\hspace{3mm}%
\begin{minipage}[c]{8.2cm}%
\centering
{\resizebox{\linewidth}{!}{%
\begin{tikzpicture}[
  font=\scriptsize\ttfamily,
  >={Stealth[length=1.6mm]},
  box/.style={draw,line width=0.5pt,rounded corners=2.5pt,align=center,
              inner xsep=5pt,inner ysep=3.5pt},
  lav/.style={box,fill=exlav,draw=exlavdark},
  grn/.style={box,fill=exgreen,draw=exgreendark},
  gry/.style={box,fill=exgray,draw=exgraydark,text=black!65},
  tag/.style={font=\scriptsize\ttfamily,inner sep=1.5pt,rounded corners=1pt,text=white},
  cnt/.style={font=\scriptsize\rmfamily,inner sep=1pt},
]
\node[lav] (c) {module->create\_loc\_conf(cf)};
\node[anchor=east,inner sep=0pt,xshift=-1mm] at (c.west) {\color{exstar}$\bigstar$};
\node[tag,fill=exgraydark,anchor=west,xshift=1.5mm] at (c.east) {Andersen: $\varnothing$};
\node[lav,below=7mm of c] (fp) {disp(c) = ngx\_http\_module\_t.create\_loc\_conf};
\draw[->,dashed,line width=0.7pt,exlavdark] (c) -- node[tag,fill=exlavdark,right=1mm]{load, offset 48} (fp);
\node[grn,below=10mm of fp.south,anchor=north] (gl)
  {create\_loc\_conf\\gzip\_create\_conf\\{\rmfamily\itshape(25 functions)}};
\node[gry,left=2mm of gl.west,anchor=east] (gs)
  {create\_srv\_conf\\upstream\_hash\_create\_conf\\{\rmfamily\itshape(6 functions)}};
\node[gry,right=2mm of gl.east,anchor=west] (gm)
  {create\_main\_conf\\map\_create\_conf\\{\rmfamily\itshape(10 functions)}};
\draw[->,dashed,line width=0.9pt,exgreendark] (gl.north) -- node[tag,fill=exgreendark,right=0.5mm]{Add-Witness} (fp.south);
\draw[->,dashed,line width=0.7pt,exgraydark] (gs.north) to[out=90,in=200] node[tag,fill=exgraydark,pos=0.35,left=0.3mm]{Prune-Store} ([xshift=-12mm]fp.south);
\draw[->,dashed,line width=0.7pt,exgraydark] (gm.north) to[out=90,in=-20] node[tag,fill=exgraydark,pos=0.35,right=0.3mm]{Prune-Store} ([xshift=12mm]fp.south);
\begin{scope}[on background layer]
\node[draw=black!45,dotted,line width=0.5pt,rounded corners=3pt,inner sep=3pt,fit=(gs)(gl)(gm)] (all) {};
\end{scope}
\node[tag,fill=black!45,anchor=north] at (all.south) {FLTA: same signature, 41};
\end{tikzpicture}}}%
\par
{\footnotesize\textbf{(b)} The dispatch relation reconstructed by \textsc{Facet}}
\end{minipage}%
\end{lrbox}
\begin{figure}[t]
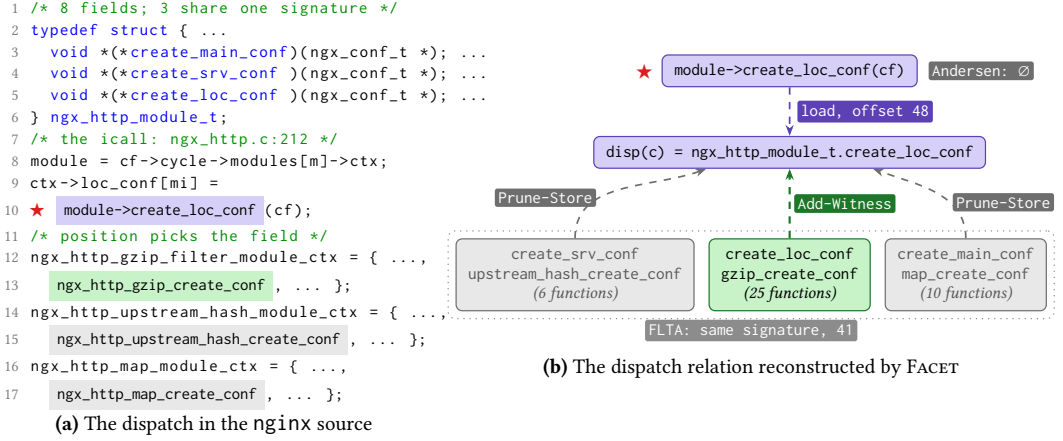

\centering
\resizebox{\linewidth}{!}{\usebox{\examplefigbox}}
\Description{(a) nginx source: the struct, three of whose eight function-pointer fields share one signature, the indirect call that loads create\_loc\_conf, and three module tables whose functions gzip\_create\_conf, upstream\_hash\_create\_conf, and map\_create\_conf are stored into three different fields by position. (b) The reconstructed relation: the call site, the dispatched field recovered from the debug layout at offset 48, and one box per field with the functions the module tables store into it (25, 6, and 10); the 25 of create\_loc\_conf are added under Add-Witness, the 16 of the sibling fields are refuted under Prune-Store, and FLTA keeps all 41 because they share one signature.}
\caption{The running example, \code{nginx}'s \code{create\_loc\_conf} dispatch.}
\label{fig:example}
\end{figure}

\subsection{The Missing Key in Existing Analyses}\label{sec:gap}

We continue with the running example, in which the dispatch is a store-to-load chain: the \emph{module table} stores the function into the field and the site loads it back.
None of the three families reconstructs this chain from the IR it is given.
\textbf{Pointer analysis} such as {Andersen's~\cite{andersen1994program}} follows value flows.
However, the flow into the field passes through the module array, which Andersen's analysis cannot traverse.
{Hence, Andersen's analysis returns the empty set, the $\varnothing$ at the site in \Cref{fig:example}(b).}
\textbf{Type-based analysis} keys the site to functions of a compatible type, but opaque IR leaves it only the signature.
Field-type matching by \mlta~\cite{lu2019where} needs the declared type of the loaded field, so it resolves none of \code{nginx}'s 351 \icalls.
Signature matching by \flta\ uses the signature that the three sibling fields share.
{Hence, it keeps all 41 functions that the \emph{module tables} store into the three fields, 25, 6, and 10, the dotted frame of \Cref{fig:example}(b).}
TypeCopilot~\cite{zhou2025typecopilot} first re-infers the erased types, which restores the matching key.
However, a recovered type set only bounds what a field may hold and does not determine which functions the program assigns to it (\S\ref{sec:eval:comparison}).
\textbf{Learning-based analysis} such as \sea~\cite{cheng2024sea} uses the source names but has no representation of the dispatched field.
At this site, \sea\ returns 42 targets: the 25 observed ones and 17 others.
Its decision is adopted without a symbolic bound or a record of evidence, so a misprediction remains uncorrected~\cite{venkatesh2024emergence,venkatesh2025empirical}.
None of them recovers the field the site dispatches through, which the IR no longer names.

The dispatch, however, is still recoverable from two facts the IR keeps.
{For the first fact, the debug information records \code{create\_loc\_conf} at the load's byte offset, 48; this is the dispatched field $\mathit{disp}(c)$ of \Cref{fig:example}(b).}
For the second fact, function constants keep their types, so the initializer slots that assign them stay identifiable; \code{nginx} has 804 such slots.
{Among these slots, the \emph{module tables} store 25 functions into \code{create\_loc\_conf}, 6 into \code{create\_srv\_conf}, and 10 into \code{create\_main\_conf}, the three boxes of \Cref{fig:example}(b); each box is the initializer set of its field.}
{Joining the two facts on the field needs no value flow: each of the 25 functions is witnessed by a \emph{module table}, and the 16 candidates of the sibling fields are refuted by the same writes, the two kinds of arrows in \Cref{fig:example}(b).}

\subsection{Dispatch Relation}\label{sec:metric}

Let $M$ be a whole-program opaque-pointer module with its debug information.
A \emph{resolver} computes a call graph $R$ that maps each \icall\ site $c$ of $M$ to a set of target functions $R(c)$.
An existing analysis, such as Andersen's, computes the initial call graph $R_0$.
Let $T(c)$, the \emph{runtime targets} of $c$, be the functions that $c$ invokes in some execution of $M$.
A call graph $R$ is \emph{sound} at $c$ if $T(c)\subseteq R(c)$.

\begin{definition}[Dispatch relation]\label{def:dispatch}
Let an \icall\ site $c$ load its callee from field $m$ of struct $S$ or from a global variable $g$.
Its \emph{field key} $\mathit{disp}(c)$ is $S{.}m$ or $g$, respectively.
Let $T(S{.}m)$, the \emph{runtime assignments} of $S{.}m$, be the functions that some execution of $M$ stores into field $m$ of an object of type $S$.
The \emph{dispatch relation} of $M$ is the set of pairs $(c,f)$ with $f\in T(\mathit{disp}(c))$.
\end{definition}

Below, $g$ is treated as $S{.}m$.
A function enters $T(S{.}m)$ through an initializer, a runtime store, or an aggregate copy.
Let the \emph{initializer set} $\mathit{init}(S{.}m)\subseteq T(S{.}m)$ be the functions that global initializers store there.
A static analysis approximates $T(S{.}m)$ by an \emph{assignment set} $F(S{.}m)$ for each field.
Hence, it approximates the dispatch relation by $F(\mathit{disp}(c))$ at each site.
Our approximation is \emph{field-based}~\cite{moller2026spa}.
It merges all objects of type $S$ into one set $F(S{.}m)$.
Let $\mathit{sig}(S{.}m)$ and $\mathit{sig}(f)$ be the signatures that the debug information declares for the field and for a function $f$.
Two signatures are \emph{compatible} if they have the same return type and arity and each pair of parameters is equal or includes \code{void*}.
We write $\mathit{sig}(f)\sim\mathit{sig}(S{.}m)$ for compatible signatures.

\section{Our Approach: Dispatch-Relation Reconstruction and Call-Graph Refinement}\label{sec:framework}
\begin{figure}[t]
\centering
\includegraphics[width=\linewidth]{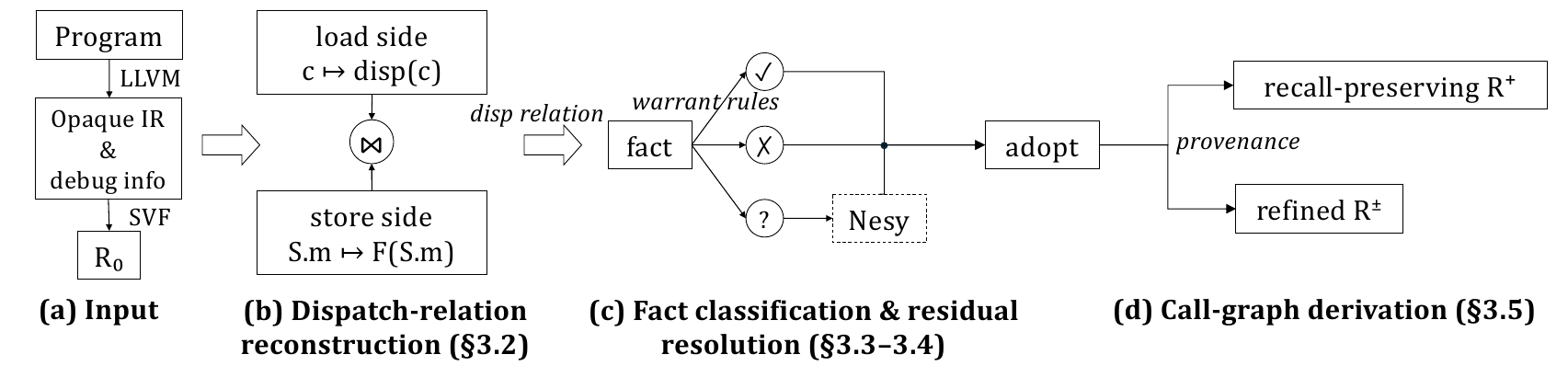}%
\Description{Overview of Facet. (a) Input: the program, compiled by LLVM to opaque IR with debug information, and the initial call graph R0 from SVF. (b) Dispatch-relation reconstruction: the load side maps each site to its dispatched field and the store side maps each field to its assigned functions; the two are joined on the field key into the dispatch relation. (c) Fact classification and residual resolution: each fact is classified by the warrant rules as supported, rejected, or residual; residual facts go to the LLM; adopted facts form the provenance. (d) Call-graph derivation: the provenance yields the recall-preserving call graph and the refined call graph.}
\caption{Overview of \ourtool.}
\label{fig:overview}
\end{figure}

\subsection{Overview}\label{sec:model}

As shown in~\Cref{fig:overview}, \ourtool\ refines $R_0$ in three stages and a final derivation:
\emph{Dispatch-relation reconstruction} (\S\ref{sec:prune}) recovers the approximate dispatch relation from the debug layouts and the points-to solution.
It proposes a \emph{call-target fact} $\langle c,f,\add\rangle$ for each $f$ in $F(\mathit{disp}(c))$.
It also proposes $\langle c,f,\prune\rangle$ for each target of $R_0(c)$ that the dispatched field excludes.
\emph{Warrant-based classification} (\S\ref{sec:verify}) sorts every fact by program evidence into \verified, \rejected, or \heuristic.
\emph{Neuro-symbolic residual resolution} (\S\ref{sec:add}) then lets an LLM choose the targets at the residual cases from candidates that the symbolic stages have bounded.
The decisions of these three stages form the provenance.
From it, \emph{call-graph derivation} (\S\ref{sec:guarantee}) builds the recall-preserving $\Rrp$ and the refined $\Rref$.

\subsection{Debug-Guided Dispatch-Relation Reconstruction}\label{sec:prune}

Because opaque IR keeps the two components of the dispatch relation in different places, we recover them separately.
The debug layouts still give the dispatched field $\mathit{disp}(c)$ of each site (\S\ref{sec:disp}).
The points-to solution still gives the assignment set $F(S{.}m)$ of each field (\S\ref{sec:proj}).
We then join the two components on the field key $S{.}m$.

\ourtool\ builds $R_0$ on Andersen's analysis.
The projection of \S\ref{sec:proj} needs the same points-to solution, so one run of the analysis serves both.
At each site $c$, $R_0(c)$ holds the functions in the points-to set of the callee pointer~\cite{moller2026spa}.
Let $\mathit{AT}$ be the address-taken functions of $M$.
Then, $R_0$ keeps only the functions in $\mathit{AT}$ whose IR function type equals the call's, as \flta\ does~\cite{tice2014forward,ge2016fine}.
In opaque IR, every pointer parameter is \code{ptr}, so this match distinguishes only the return type, the arity, and the non-pointer parameters.
We use this filter because it is the coarsest type filter that opaque IR still supports.
The filter removes only functions of another IR type, so $R_0$ keeps Andersen's recall under the typability assumption of \S\ref{sec:prunerules}.

\subsubsection{Dispatched-Field Recovery from Debug Layouts}\label{sec:disp}

At an \icall\ $c$, the callee is loaded from a base pointer $p$ at a constant byte offset $k$.
In optimized opaque IR, a field access is byte arithmetic and names no struct.
Hence, we compute $k$ by adding the byte offsets along the chain of address computations before the load.
To recover $\mathit{disp}(c)=S{.}m$, we need the struct type $S$ that $p$ points to.
The field $m$ is then the field at offset $k$ in the layout of $S$.
Neither $S$ nor its layout is in the IR.
However, the debug information still records source-level types~\cite{hueck2025dimeta}.
Therefore, we recover both from the recorded type of $p$.

To find that type, \ourtool\ tries three steps in order and stops at the first that succeeds.
\ding{182}~\emph{Direct.} A debug record binds $p$ to a source variable whose recorded type is a pointer to $S$.
In optimized IR, this is a value record on $p$ itself.
In unoptimized IR, it is a declaration record on the stack slot from which $p$ is loaded.
After inlining, $p$ also has records for locations derived from it, such as a lock inside the object.
The types of these locations are not the type of $p$.
Only a record for $p$ itself has an empty location expression, because a record for a derived location expresses an offset from $p$.
Hence, we accept only such records.
\ding{183}~\emph{Chained.} Step~\ding{182} fails when no variable names $p$.
This step applies when $p$ is loaded from a field of a receiver object.
For example, the inner pointer in \code{cinfo->master->prepare()} is a temporary loaded from the field \code{master} of \code{cinfo}.
We resolve the receiver by the same steps.
We then follow the recorded type of that field to the pointee struct.
To bound this recursion, we cap its depth at four.
The depth counts the pointers from the resolved receiver to $p$, two in the example.
A chain beyond the cap is left unrecovered.
\ding{184}~\emph{Typed access.} In unoptimized IR, the address computation still names a struct type.
We look this name up in the debug information.
{Because steps~\ding{182} and~\ding{183} use only the debug layout and the load's byte offset, the recovery requires only debug information in the bitcode, at any optimization level.}

Once $S$ is fixed, three cases need a different field key, because the plain layout lookup at offset $k$ would split or miss the writes to one location.
First, when the field that covers offset $k$ is an embedded named struct, we descend into it.
Hence, a wrapper type that dispatches through an embedded base struct is keyed on the base struct's field.
Its sites then join the writes made through the base type, as \code{my\_upsampler} does through \code{jpeg\_upsampler}.
Second, an anonymous union is one location, so the field key stops at the enclosing named field and covers every member.
Third, a dispatch through a plain global function pointer or a constant-indexed global array involves no struct, so we key it on the variable name.
The site of \Cref{fig:example} needs no adjustment: the callee is loaded at offset 48 from a pointer recorded as \code{ngx\_http\_module\_t*}, and the layout places \code{create\_loc\_conf} there with signature \code{void*(ngx\_conf\_t*)}.
This declared signature is $\mathit{sig}(S{.}m)$, the type evidence the projection's filter and \textsc{Prune-Sig} (\S\ref{sec:prunerules}) use.

When none of the three steps applies, the symbolic stages yield no fact at the site.
The recovery fails when the receiver is named by no debug record and reached by no typed chain, when the base pointer is copied into a local of another type, or when the index is not constant.
Such sites are $10.2\%$ of the ground-truth sites (\S\ref{sec:eval:recon}).
The residual resolution of \S\ref{sec:add} still applies there.

\subsubsection{Field-Assignment Recovery by Points-to Projection}\label{sec:proj}
\begin{figure}[]
\centering
%
\definecolor{addgreen}{RGB}{0,140,0}
\definecolor{keyfill}{RGB}{226,238,247}
\begin{tikzpicture}[
  x=1mm,y=1mm,
  font=\scriptsize,
  >={Stealth[length=1.5mm,width=1.2mm]},
  ctitle/.style={font=\scriptsize\bfseries,inner sep=0pt,anchor=south},
  rel/.style={inner sep=0pt,font=\scriptsize},
  op/.style={draw=black!75,line width=0.5pt,fill=white,inner xsep=2.2pt,inner ysep=1.6pt,align=center,font=\scriptsize,minimum width=27mm},
  lab/.style={font=\tiny,inner sep=0.8pt},
  arr/.style={->,line width=0.6pt,draw=black!75},
  joinnode/.style={circle,draw=black!80,line width=0.6pt,fill=white,inner sep=0pt,minimum size=4.2mm,font=\footnotesize},
]
\renewcommand{\arraystretch}{0.9}\setlength{\tabcolsep}{2pt}
\newcommand{\hk}{\cellcolor{keyfill}}
\node[ctitle,anchor=east] at (9.5,21) {load side};
\node[ctitle,anchor=east] at (9.5,7)  {store side};
\node[rel,anchor=east] (L) at (49,21) {\begin{tabular}{|c|c|}\hline
 site & load \\ \hline
 $c$ & \texttt{module->create\_loc\_conf} \\ \hline
\end{tabular}};
\node[op,anchor=west] (rec) at (52,21) {recover the field (\S\ref{sec:disp})};
\node[rel,anchor=east] (C) at (106,21) {\begin{tabular}{|c|c|}\hline
 site & $\mathit{disp}(c)$ \\ \hline
 $c$ & \hk$S{.}m$ \\ \hline
\end{tabular}};
\draw[arr] (L.east) -- (rec.west);
\draw[arr] (rec.east) -- (C.west);
\node[rel,anchor=east] (W) at (49,7) {\begin{tabular}{|c|c|c|}\hline
 $o$ & field & $f$ \\ \hline
 $o_1$ & \hk\texttt{create\_loc\_conf} & \textcolor{addgreen}{$f_1$} \\
 $\vdots$ & \hk$\vdots$ & $\vdots$ \\
 $o_{25}$ & \hk\texttt{create\_loc\_conf} & \textcolor{addgreen}{$f_{25}$} \\ \hline
\end{tabular}};
\node[op,anchor=west] (proj) at (52,7) {project onto the field key $S{.}m$};
\node[rel,anchor=east] (F) at (106,7) {\begin{tabular}{|c|c|}\hline
 field & $F(S{.}m)$ \\ \hline
 \hk$S{.}m$ & $\{\textcolor{addgreen}{f_1,\ldots,f_{25}}\}$ \\ \hline
\end{tabular}};
\draw[arr] (W.east) -- (proj.west);
\draw[arr] (proj.east) -- (F.west);
\node[joinnode] (join) at (113,14) {$\bowtie$};
\draw[arr] (C.east) -| (join.north);
\draw[arr] (F.east) -| (join.south);
\node[rel,anchor=west] (R) at (121,14) {\begin{tabular}{|c|}\hline
 $F(\mathit{disp}(c))$ \\ \hline
 $\{\textcolor{addgreen}{f_1,\ldots,f_{25}}\}$ \\ \hline
\end{tabular}};
\draw[arr] (join.east) -- (R.west);
\end{tikzpicture}%
\Description{Field-assignment projection on the running example, drawn as a query plan over relations. The store side is the relation Writes(o, S.m, f): 25 objects of layout ngx_http_module_t, each with the field create_loc_conf and one function. Projecting onto the key (S, m) yields F(S.m), the set of the 25 functions. The load side is the relation Sites(c, disp(c)) with the one site and the same field. Joining F and Sites on the field yields the candidates F(disp(c)) at c, the same 25 functions.}
\caption{Field-assignment projection on the running example (\code{create\_loc\_conf} at \code{ngx\_http.c:212}).}
\label{fig:projection}
\end{figure}
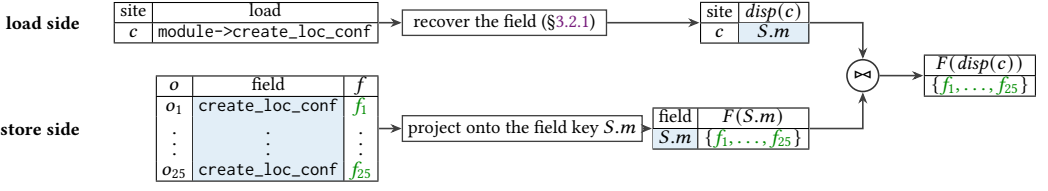

\begin{wrapfigure}{r}{0.5\linewidth}
\footnotesize
\hrule
\captionof{algorithm}{Assignment reconstruction.}\label{alg:proj}
\hrule
\begin{algorithmic}[1]
\algtext*{EndFor}\algtext*{EndIf}
\Require module $M$, points-to solution $\mathit{pt}$, debug layouts
\Ensure $\mathit{init}(S{.}m)\subseteq F(S{.}m)\subseteq \mathit{AT}$
\Statex $\mathit{key}(p)$: the field $S{.}m$ that address $p$ resolves to (\S\ref{sec:disp})
\Statex $\mathit{pt}(o{.}m)$: the points-to set of field $m$ of abstract object $o$
\Statex each $\mathrel{+}=$ on $F(S{.}m)$ keeps only $f$ with $\mathit{sig}(f)\sim\mathit{sig}(S{.}m)$
\State $\mathit{init}(S{.}m),F(S{.}m)\gets\varnothing$ for every function-pointer field $S{.}m$
\ForAll{initialized globals $g$} \Comment{form 0}
  \ForAll{function-pointer slots $S{.}m$ of $g$ holding $f$}
    \State $\mathit{init}(S{.}m)\mathrel{+}=f$;\ $F(S{.}m)\mathrel{+}=f$
  \EndFor
\EndFor
\ForAll{$\mathtt{store}\ v,p$, $\mathit{key}(p)=S{.}m$} \Comment{form 1}
  \State $F(S{.}m)\mathrel{+}=\mathit{pt}(v)$
\EndFor
\ForAll{$\mathtt{memcpy}(d,s,n)$, $d$ resolves to type $S$} \Comment{form 2}
  \ForAll{function-pointer fields $m'$ of $S$}
    \State $F(S{.}m')\mathrel{+}=\bigcup_{o\in\mathit{pt}(s)}\mathit{pt}(o{.}m')$
  \EndFor
\EndFor
\State \Return $\mathit{init},F$
\end{algorithmic}
\hrule
\end{wrapfigure}
To compute the assignment set $F(S{.}m)$ of \S\ref{sec:metric}, \ourtool\ projects the points-to solution onto fields (\Cref{alg:proj}).
To project is to index each write by the field key of its address rather than by the object it writes.
Hence, every store into field $m$ of any object of type $S$ contributes to $F(S{.}m)$.
We choose this projection because it gives the assignment set the same key as the dispatched field.
Thus, the two components can be joined directly.
Merging the objects of one type is an established heap abstraction~\cite{tan2017mahjong}, but it loses precision where they hold different targets (\S\ref{sec:eval:comparison}).

Formally, let $\mathit{pt}(o{.}m)$ be the points-to set of field $m$ of abstract object $o$ in the field-sensitive points-to solution $\mathit{pt}$~\cite{moller2026spa}, and $\mathit{type}(o)$ the type of $o$.
Then $F(S{.}m)$ holds the functions $f\in\bigcup_{\mathit{type}(o)=S}\mathit{pt}(o{.}m)$ with $\mathit{sig}(f)\sim\mathit{sig}(S{.}m)$.
Opaque IR records no $\mathit{type}(o)$ for a heap object, so \Cref{alg:proj} keys each write by the recorded type of its address instead, through the recovery of \S\ref{sec:disp}.
Because a function reaches $\mathit{pt}(o{.}m)$ through three kinds of statements, \Cref{alg:proj} handles them in three forms.
Form~0 traverses every global constant initializer under the global's recorded type and records each function that the initializer places in a field.
We keep this set separately as $\mathit{init}(S{.}m)$ for the witness of \S\ref{sec:addrules}.
\Cref{fig:projection} shows Form~0 on the running example as a query plan.
The 25 \emph{module tables} each store one function into the field \code{create\_loc\_conf}.
Projecting these writes onto the field key $S{.}m$ yields $F(S{.}m)$, and joining it with the site's recovered field yields the site's 25 candidates.

Forms~1 and~2 cover the writes that are not constant initializers.
A function is often copied or passed as a parameter before it is stored into a field.
Form~1 handles these stores.
It keys the store address by the recovery of \S\ref{sec:disp}.
Because counting only stores of a function constant would miss most assignments (\S\ref{sec:eval:recon}), it takes the stored values from the points-to set that Andersen's analysis has already computed.
An aggregate assignment lowers to a bulk copy with no store instruction.
Form~2 handles these copies.
It distributes the copy field by field across the copied type.
Because the projection widens $F$, we apply it only to fields that the debug information types as function pointers.
We also keep only the signature-compatible candidates.
The two components are then joined on the field key, so the product of reconstruction at a site $c$ is the candidate set $F(\mathit{disp}(c))$.
Where the field is unrecovered or the assignment set is empty, reconstruction supplies no evidence, and \S\ref{sec:add} resolves these residual cases.

\subsection{Warrant-Based Fact Classification}\label{sec:verify}

Warrant-based classification decides which proposals may modify the graph.
The rules in \Cref{fig:rules} classify every fact against program evidence.
A classification is written as $\langle c,f,\add\rangle\Downarrow v$ or $\langle c,f,\prune\rangle\Downarrow v$ under the rule named beside it, and its outcome $v$ is \verified\ or \rejected.
For an \add, \verified\ means that the edge is added, and \rejected\ means that it is withheld.
For a \prune, \verified\ means that the edge is removed.
In every rule, $S{.}m$ is the dispatched field $\mathit{disp}(c)$.
{The rules take the recovered field as $\mathit{disp}(c)$.}
{Its agreement with typed IR is measured in \S\ref{sec:eval:recon}.}
The refutations are tried before the witnesses, and the first rule that applies decides.
A fact that no rule classifies is \heuristic.
Each \verified\ or \rejected\ fact carries a warrant.
\begin{definition}[Warrant]\label{def:warrant}
A \emph{warrant} for a call-target fact is a triple $\langle r,e,\pi\rangle$.
Here, $r$ is the rule of \Cref{fig:rules} that classifies the fact.
The evidence $e$ is what $r$ checked.
The premise $\pi$ is an assumption about $M$ under which $e$ justifies the decision.
It is \emph{premise-free} if $\pi{=}\mathit{true}$ and \emph{premised} otherwise.
\end{definition}

\noindent \ourtool\ checks every condition above the bar of a rule, except the premise $\pi$.
It assumes $\pi$ and records it with the decision.
\ourtool\ cannot confirm a premise from the IR alone, because a premise concerns what the program does at run time.
As in argumentation theory~\cite{toulmin1958uses}, a warrant justifies one edge decision under $\pi$ and makes no claim about whether the edge is ever taken.
For a supported \add, it justifies that the program assigns $f$ to the dispatched field, $f\in T(S{.}m)$ of \Cref{def:dispatch}.
For a rejected \add\ or a supported \prune, it justifies that $f$ is not a target of $c$, $f\notin T(c)$.

Among the rules, only \textsc{Add-Witness} is premise-free.
The other five rules are premised, and \Cref{fig:rules} states each premise.
The premise-free rule uses only the initializers.
Each premised rule uses an approximation instead.
Its premise states what the approximation must get right for the decision to hold.
A premise-free decision is justified unconditionally, and a premised one only where $\pi$ holds.
A soundy analysis must name that premise for its client~\cite{livshits2015soundiness}, and \S\ref{sec:guarantee} derives two call graphs from the distinction.
\S\ref{sec:eval:attribution} measures how often the premises of the prune rules fail.
The rules implement the asymmetry of \textbf{C2}: an addition needs evidence that the target is possible, whereas a removal needs evidence that \emph{excludes} it under a stated assumption.

\begin{figure}[t]
\scriptsize
{\footnotesize\textit{(a) Witnesses (\S\ref{sec:addrules})}}
\begin{mathpar}
\mprset{lineskip=0.35em,sep=1.5em,andskip=0.5em plus 0.5fil minus 1em}
\inferrule*[left={\scriptsize[\textsc{Add-Witness}]}]{f\in\mathit{init}(S{.}m)}
  {\langle c,f,\add\rangle \Downarrow \verified}
\and
\inferrule*[left={\scriptsize[\textsc{Add-Store}]}]{f\in F(S{.}m)\setminus\mathit{init}(S{.}m) \wedge \pi_{\mathrm{bind}}}
  {\langle c,f,\add\rangle \Downarrow \verified}
\end{mathpar}
{\footnotesize\textit{(b) Refutations (\S\ref{sec:prunerules})}}
\begin{mathpar}
\mprset{lineskip=0.35em,sep=1.5em,andskip=0.5em plus 0.5fil minus 1em}
\inferrule*[left={\scriptsize[\textsc{Prune-Store}]}]{F(S{.}m)\neq\varnothing \wedge f\notin F(S{.}m) \wedge \pi_{\mathrm{proj}}}
  {\langle c,f,\prune\rangle \Downarrow \verified}
\and
\inferrule*[left={\scriptsize[\textsc{Add-Unstored}]}]{F(S{.}m)\neq\varnothing \wedge f\notin F(S{.}m) \wedge \pi_{\mathrm{proj}}}
  {\langle c,f,\add\rangle \Downarrow \rejected}
\\
\inferrule*[left={\scriptsize[\textsc{Prune-Sig}]}]{\mathit{sig}(f)\not\sim\mathit{sig}(S{.}m) \wedge \pi_{\mathrm{sig}}}
  {\langle c,f,\prune\rangle \Downarrow \verified}
\and
\inferrule*[left={\scriptsize[\textsc{Add-Sig}]}]{\mathit{sig}(f)\not\sim\mathit{sig}(S{.}m) \wedge \pi_{\mathrm{sig}}}
  {\langle c,f,\add\rangle \Downarrow \rejected}
\end{mathpar}
\caption{Inference rules for call-target facts.}
\label{fig:rules}
\Description{Inference rules in two groups: (a) the witnesses Add-Witness and Add-Store, which support an add; (b) the refutations, each removal rule beside the add rejection with the same premise (Prune-Store with Add-Unstored, Prune-Sig with Add-Sig).}
\end{figure}

\subsubsection{Warrants for Additions}\label{sec:addrules}

As \Cref{fig:rules}(a) shows, an \add\ is supported only by a positive witness that the program supplies.
The first witness is \textsc{Add-Witness}, the \emph{initializer witness} and the strongest rule.
It applies when $f$ is in the initializer set $\mathit{init}(S{.}m)$ for $S{.}m=\mathit{disp}(c)$.
An initializer is program text, so $f\in\mathit{init}(S{.}m)\subseteq T(S{.}m)$ holds by the definition of $\mathit{init}$ without assumption.
In the running example, it warrants the 25 functions of \code{create\_loc\_conf}.
The second witness is \textsc{Add-Store}, the \emph{runtime-store witness}.
It applies when $f\in F(S{.}m)\setminus\mathit{init}(S{.}m)$.
In this case, a store or a copy brings $f$ into the field, but no initializer does.
{Its evidence is weaker than that of the initializer witness, which names $f$ in the field.}
In contrast, a store contributes only the may points-to set of its value, and a copy contributes the points-to sets of the objects its source may point to.
Hence, the witness needs the premise $\pi_{\mathrm{bind}}$ that the projected assignment is realized: $f\in T(S{.}m)$.

\subsubsection{Warrants for Removals}\label{sec:prunerules}

As \Cref{fig:rules}(b) shows, a \prune\ is supported only by evidence that excludes the target.
Such evidence must come from the \emph{complement of a sound over-approximation}: if a relation contains every function that may be written into a field, a function outside it is refuted~\cite{moller2026spa}.
The complement of a relation that may have missed members refutes nothing, so an empty $F(S{.}m)$ is no evidence of absence.
Two rules satisfy this requirement without consulting the model.

\textsc{Prune-Store} refutes $f$ when $F(S{.}m)$ is non-empty and omits $f$.
Its premise $\pi_{\mathrm{proj}}$ has two parts.
{First, the recovered field key $S{.}m$ is the field $c$ loads from.}
{Second, the projection is sound at that field: it misses no assignment, so $T(S{.}m)\subseteq F(S{.}m)$.}
\textsc{Prune-Sig} refutes $f$ when its signature is not compatible with the field's, $\mathit{sig}(f)\not\sim\mathit{sig}(S{.}m)$.
Its premise $\pi_{\mathrm{sig}}$ is well-typed dispatch: every target of $c$ has a signature compatible with the field's.
This is the typability assumption under which a points-to set may be filtered by type.
By the definition of $\sim$, the refutation applies only to a return-type, arity, or concrete-pointer mismatch.
In the running example, the \emph{module tables} never store the functions of \code{create\_srv\_conf} and \code{create\_main\_conf} into \code{create\_loc\_conf}.
As a result, \textsc{Prune-Store} refutes all of them.
In contrast, the three sibling fields share one signature, so \textsc{Prune-Sig} separates none of their functions.

The same refutations also reject an \add.
\textsc{Add-Unstored} and \textsc{Add-Sig} reject the \add\ of any target that \textsc{Prune-Store} or \textsc{Prune-Sig} refutes.
They use the same premises as these two rules.
They are tried before the witnesses, so the rules never support an \add\ and a \prune\ of the same edge.
Because the model names no over-approximation, a model-proposed \prune\ that neither rule supports stays \heuristic.
The cases that no rule decides are left to \S\ref{sec:add}.

\subsection{Neuro-Symbolic Residual Resolution}\label{sec:add}

The rules of \S\ref{sec:verify} leave two kinds of \emph{residual cases}: sites that remain empty after the witnessed additions, and candidates of the same signature that no warrant separates.
In both kinds, the candidates can still differ in their semantic role.
The type and points-to facts that \ourtool\ recovers do not express this role.
{However, the names and implementations of the candidates do.}
{Hence, \ourtool\ resolves the residual cases via a neuro-symbolic approach:}
{Before a query, the symbolic stages bound the candidates (\S\ref{sec:queries}).}
{The LLM then decides which of these candidates are targets, which is the only neural step.}
{After the query, the symbolic rules check each decision and determine whether it is adopted (\S\ref{sec:adopt}).}

\subsubsection{Completion and Disambiguation}\label{sec:queries}

A prune-only resolver cannot exceed the recall of its input, so no filter can populate an initially empty site.
Hence, for a site the witnesses leave empty, \emph{completion} lets the model select \add\ targets.
The symbolic stages bound its candidates to the address-taken functions of the site's signature category.
{When recovered, the field serves as context rather than evidence: the query names it and asks for \emph{this} field's implementors so that the model can distinguish similarly named siblings.}
\emph{Disambiguation} applies at the same-signature residual cases, where the candidate set is already symbolically bounded.
Hence, a single prompt per site presents every candidate by its name and body, and asks for a keep-or-prune decision on each.
{We present the candidates together so that the model can \emph{compare} siblings rather than classify each in isolation.}
{In both queries, the model answers yes or no for each listed candidate and therefore cannot name a function outside the bound.}

\subsubsection{Warrant Precedence and Abstention}\label{sec:adopt}

The rules of \S\ref{sec:verify} classify each model decision like any other fact.
For example, \textsc{Add-Sig} rejects a decision whose target has a signature that the dispatched field excludes.
Hence, a model decision receives a warrant only where program evidence supports it.
When the model gives no usable answer, the symbolic result is kept.

The refined call graph $\Rref$ adopts residual facts and every warranted fact, including the premised ones, subject to two checks.
Under \emph{precedence}, the warranted decision is kept when a warranted rule and a model proposal disagree on an edge.
Hence, a residual \prune\ acts only on edges that no warrant supports.
Under \emph{abstention}, whether a change is adopted at all depends on the site, for warranted and residual changes alike.
At an initially empty site, an addition cannot lower recall.
At a site that the evidence has already narrowed to one or two targets, however, a change in either direction is more likely to harm than to help (\S\ref{sec:eval:attribution}).
The \emph{symbolic} result at a site is $R_0$ after the refutations of \S\ref{sec:prunerules}.
Therefore, \ourtool\ abstains wherever the symbolic result is non-empty and holds at most $\tau$ targets.
We call $\tau$ the abstention threshold.
Let $P_{\mathrm{sym}}(c)$ be the targets that \textsc{Prune-Store} and \textsc{Prune-Sig} refute at $c$.
Then \ourtool\ abstains at $c$ iff $0<|R_0(c)\setminus P_{\mathrm{sym}}(c)|\le\tau$.
Such a site is returned as the symbolic analysis left it.
That is, no addition is adopted there, including warranted ones.
The model's removals are withdrawn, but warranted refutations stay.
Changes apply to a copy of $R_0$, so classification does not depend on the order in which facts are adopted.

\subsection{Call-Graph Derivation from the Provenance}\label{sec:guarantee}

To produce both call graphs from a single analysis, \ourtool\ runs the stages once, in a fixed order.
\ding{182}~Build $R_0$.
\ding{183}~Recover $\mathit{disp}(c)$ at every site, and $\mathit{init}$ and $F$ for every field (\Cref{alg:proj}).
\ding{184}~Propose an \add\ for each $f$ in the candidate set $F(\mathit{disp}(c))$, and a \prune\ for each target of $R_0(c)$ that \textsc{Prune-Store} or \textsc{Prune-Sig} refutes.
Classify these facts by the rules of \Cref{fig:rules}.
\ding{185}~Form the symbolic result at each site.
From it, find the sites where \ourtool\ abstains.
\ding{186}~Query the model only at the residual cases: for completion at sites that are still empty, and for disambiguation at the same-signature residual cases.
\ding{187}~Adopt under precedence and abstention.

Each stage records what it decides.
A \emph{decision} is an \add\ or a \prune\ of $f$ at $c$, together with its outcome.
A \verified\ or \rejected\ decision also records the warrant $\langle r,e,\pi\rangle$ that $r$ supplies (\Cref{def:warrant}).
The \emph{provenance} of a run is the set of its decisions~\cite{zhao2020provenance,raghothaman2020prosynth}.
Both call graphs are derived from the provenance without re-running the analysis.
Because the runtime targets cannot be observed from $M$ alone, their guarantees are stated relative to $R_0$.
The \emph{recall-preserving call graph} $\Rrp$ adds only edges that premise-free decisions support.
Because no \prune\ rule is premise-free, it keeps every edge of $R_0$, and its additions are those of \textsc{Add-Witness} at sites where \ourtool\ does not abstain.
A rejected \add\ only withholds an edge, so $\Rrp$ honors it although it is premised.
Formally, $A_W(c)$, $A_S(c)$, and $A_{\mathrm{res}}(c)$ are the targets whose \add\ at $c$ is supported by \textsc{Add-Witness}, supported by \textsc{Add-Store}, or \heuristic, respectively.
Likewise, $P_{\mathrm{res}}(c)$ are the targets whose \prune\ at $c$ is \heuristic.
As precedence requires, these targets carry no add warrant.
The refuted targets $P_{\mathrm{sym}}(c)$ are as in \S\ref{sec:adopt}.
Finally, $[X]_c$ is $\varnothing$ if \ourtool\ abstains at $c$ and $X$ otherwise, so that an abstaining site receives no addition and no residual removal:
\begin{align*}
\Rrp(c) &= R_0(c)\ \cup\ [A_W(c)]_c,\\
\Rref(c) &= \bigl(R_0(c)\setminus(P_{\mathrm{sym}}(c)\cup[P_{\mathrm{res}}(c)]_c)\bigr)\ \cup\ [A_W(c)\cup A_S(c)\cup A_{\mathrm{res}}(c)]_c.
\end{align*}

Hence, $R_0(c)\subseteq\Rrp(c)$ at every site, and $\Rrp$ never falls below the recall of $R_0$.
The \emph{refined call graph} $\Rref$ forgoes this guarantee to admit premised pruning and residual inference.
Following defensive points-to analysis~\cite{smaragdakis2018defensive}, \ourtool\ records the premise behind every change (\S\ref{sec:verify}).
{Hence, single analysis yields two call graphs.}
{The recall-preserving $\Rrp$ serves clients that must not lose an initial edge, such as forward-edge CFI~\cite{abadi2005cfi,tice2014forward,burow2017cfi}.}
{The refined $\Rref$ serves clients that accept a small recall risk for a smaller graph, such as bug detection~\cite{sui2012saber,shi2018pinpoint,lu2019crix}.}

\section{Implementation}\label{sec:impl}

\ourtool\ consists of over 3{,}100 lines of C++ on \svf~3.3~\cite{sui2016svf} and LLVM~21.1.0 and 3{,}700 lines of Python.
Each subject is built in its default configuration with \code{-g} added.
In every configuration, we key \svf's field objects on opaque IR by base type and byte offset and count allocation wrappers as allocation sites.
{The LLM is set to Gemini~3.7~Flash at temperature~0.}
The abstention threshold $\tau$ is set to 2. To reduce randomness, we conducted three repetitions of each experiment and reported the average statistical results. {Experiments run on two Intel Xeon Gold~6248 processors and 188\,GB RAM.}

\section{Evaluation}\label{sec:eval}
\subsection{Setup}\label{sec:eval:setup}

\subsubsection{Research Questions}

Our evaluation answers five research questions:
\begin{itemize}[leftmargin=10pt,itemsep=0pt,topsep=2pt]
\item \textbf{RQ1}~(reconstruction): how much of the dispatch relation does \ourtool\ reconstruct from opaque IR, and how often does the recovered dispatched field agree with typed IR?
\item \textbf{RQ2}~(effectiveness): how much does the refined call graph improve precision and recall over $R_0$?
\item \textbf{RQ3}~(comparison): how does \ourtool\ compare with state-of-the-art \icall\ analyses?
\item \textbf{RQ4}~(ablation study): how much does each component of \ourtool\ contribute to the result?
\item \textbf{RQ5}~(application): does an audit guided by the refined call graph find real defects?
\end{itemize}

\subsubsection{Metrics}
The \emph{average number of targets per site} is $|R(c)|$ averaged over the sites that a resolver reports, which measures precision without ground truth~\cite{lu2019where,cai2024kelp}.
The \emph{Ratio} column of \Cref{tab:targets} divides the $R_0$ mean by \ourtool's.
Against the ground truth, we report per-site macro-averaged precision, recall, and F1, as \sea\ and KallGraph do~\cite{cheng2024sea,li2025kallgraph}.
Recall is taken against the targets observed at a site, a subset of its true targets.
The target average of \Cref{fig:persubject}(b) is taken over the ground-truth sites only, so that it is comparable with the ground-truth metrics.
In every table, a darker cell is better within its column.
The target columns of \Cref{tab:targets} are shaded on a logarithmic scale.

\subsubsection{Baselines}\label{sec:eval:baselines}

We compare against the aforementioned three families.
\ding{182}~\textbf{Pointer analysis}: Andersen's analysis and the initial call graph $R_0$ that it induces under \flta.
\ding{183}~\textbf{Type-based}: TypeDive~\cite{lu2019where}, the release of \mlta; KallGraph~\cite{li2025kallgraph}; TFA~\cite{liu2024tfa}, which co-analyzes types with data flow; and TypeCopilot~\cite{zhou2025typecopilot}, which re-infers the erased pointee types on opaque IR and matches on them.
\mlta\ resolves none of our \icalls\ on opaque IR (\S\ref{sec:gap}).
KallGraph and TFA were developed on typed-pointer IR~\cite{li2025kallgraph,liu2024tfa}.
Therefore, as the typed-IR control, we rebuild every subject on LLVM~14 typed-pointer IR and run all three tools there as released.\footnote{The two builds share their sites. An \icall\ exists in the LLVM~14 build at 99.3\% of the ground-truth sites (2{,}063/2{,}078). Of the 15 missing sites, 11 are in \code{libjpeg-turbo}'s TurboJPEG API files, which that build does not link.}
DeepType~\cite{xia2024deeptype} and TyPM~\cite{lu2023typm} are excluded since KallGraph's evaluation supersedes both~\cite{li2025kallgraph}.
{Kelp~\cite{cai2024kelp} is excluded because it has no public artifact.}
\ding{184}~\textbf{Learning-based}: \sea~\cite{cheng2024sea}, the state of the art in this family.

\noindent\textbf{Ablation variants.}
{Seven ablations isolate the stages, in three groups.}
{\ding{182}~\textbf{Single source}: \textsc{Prune-only}, \textsc{Witness-only}, and \textsc{Oracle-only} adopt the facts of one source alone, and \textsc{Candidates-only} keeps every signature-compatible candidate that the oracle chooses from at the initially empty sites.}
{{For example, \textsc{Witness-only} is $R_0(c)\cup A_W(c)\cup A_S(c)$, which adds the witnessed additions to the initial call graph.}}
{\ding{183}~{\textbf{No pruning rules}: \ourtool\ \emph{w/o pruning rules} sets the symbolic refutations $P_{\mathrm{sym}}$ to $\varnothing$.}}
{These five adopt every fact as classified, without precedence or abstention, so each measures what its source supplies.}
\ding{184}~\textbf{No LLM}: \textsc{Rules-only} (i.e., $R_0(c)\setminus P_{\mathrm{sym}}(c)$) and \ourtool\ \emph{w/o LLM} (i.e., the refined call graph $\Rref$ with {$A_{\mathrm{res}}=P_{\mathrm{res}}=\varnothing$}).

\subsubsection{Datasets}\label{sec:eval:gt}
\begin{table}[t]
\centering
\caption{Subjects, ground truth, and reconstruction. Darker cells are larger within a column.}
\label{tab:subjects}\label{tab:recon}
\scriptsize\renewcommand{\arraystretch}{0.95}
\renewcommand{\code}[1]{\texttt{#1}}%
\setlength{\tabcolsep}{2.5pt}
\begin{tabular}{@{}lrrrrrrrrrrrrr@{}}
\toprule
& \multicolumn{6}{c}{subject and ground truth} & \multicolumn{3}{c}{dispatched field (\%)} & \multicolumn{2}{c}{assignments} & \multicolumn{2}{c}{typed-IR check} \\
\cmidrule(lr){2-7}\cmidrule(lr){8-10}\cmidrule(lr){11-12}\cmidrule(l){13-14}
Project & Ver. & KLoC & \#\icall & GT & cov & edges & field & \makecell[r]{array,\\global} & unrec. & pairs & \makecell[r]{scan\\misses} & both & \makecell[r]{agree\\(\%)} \\
\midrule
\href{https://github.com/dovecot/core}{\code{dovecot}} & 2.3.21 & 620.1 & 1{,}347 & 269 & 20\% & 426 & \cellcolor{heat!67}94.1 & \cellcolor{heat!6}0.4 & \cellcolor{heat!11}5.6 & 2{,}478 & \cellcolor{heat!35}46.4 & 194 & \cellcolor{heat!22}92.8 \\
\href{https://www.gnu.org/software/gdbm/}{\code{gdbm}} & 1.24 & 27.2 & 32 & 24 & 75\% & 81 & \cellcolor{heat!44}62.5 & \cellcolor{heat!18}16.7 & \cellcolor{heat!36}20.8 & 101 & \cellcolor{heat!6}0.0 & 13 & \cellcolor{heat!68}100.0 \\
\href{https://github.com/libarchive/libarchive}{\code{libarchive}} & 0c353e1 & 253.8 & 241 & 76 & 32\% & 261 & \cellcolor{heat!59}82.9 & \cellcolor{heat!7}1.3 & \cellcolor{heat!28}15.8 & 394 & \cellcolor{heat!10}5.8 & 96 & \cellcolor{heat!68}100.0 \\
\href{https://github.com/davea42/libdwarf-code}{\code{libdwarf}} & 0.9.2 & 165.7 & 413 & 87 & 21\% & 96 & \cellcolor{heat!63}88.5 & \cellcolor{heat!6}0.0 & \cellcolor{heat!20}11.5 & 49 & \cellcolor{heat!10}6.1 & 148 & \cellcolor{heat!68}100.0 \\
\href{https://github.com/libjpeg-turbo/libjpeg-turbo}{\code{libjpeg-tb}} & 1db93c2 & 123.0 & 2{,}320 & 238 & \textbf{10\%} & 357 & \cellcolor{heat!67}94.5 & \cellcolor{heat!8}2.5 & \cellcolor{heat!6}2.9 & 722 & \cellcolor{heat!68}99.6 & 160 & \cellcolor{heat!68}100.0 \\
\href{https://gitlab.gnome.org/GNOME/libxml2}{\code{libxml2}} & ddcb79d & 207.1 & 1{,}957 & 170 & \textbf{9\%} & 266 & \cellcolor{heat!49}68.8 & \cellcolor{heat!9}4.7 & \cellcolor{heat!45}26.5 & 590 & \cellcolor{heat!48}67.5 & 128 & \cellcolor{heat!68}100.0 \\
\href{https://github.com/lua/lua}{\code{lua}} & 5.5 & 34.2 & 84 & 15 & 18\% & 176 & \cellcolor{heat!42}60.0 & \cellcolor{heat!6}0.0 & \cellcolor{heat!68}40.0 & 508 & \cellcolor{heat!49}69.5 & 11 & \cellcolor{heat!10}90.9 \\
\href{https://github.com/mity/md4c}{\code{md4c}} & 65c6c9d & 12.4 & 205 & 77 & 38\% & 80 & \cellcolor{heat!55}77.9 & \cellcolor{heat!6}0.0 & \cellcolor{heat!38}22.1 & 11 & \cellcolor{heat!34}45.5 & 127 & \cellcolor{heat!68}100.0 \\
\href{https://github.com/nginx/nginx}{\code{nginx}} & 1.31.3 & 249.9 & 351 & 120 & 34\% & 425 & \cellcolor{heat!37}51.7 & \cellcolor{heat!23}23.3 & \cellcolor{heat!43}25.0 & 1{,}168 & \cellcolor{heat!47}65.8 & 45 & \cellcolor{heat!68}100.0 \\
\href{https://github.com/kkos/oniguruma}{\code{oniguruma}} & 6.9.9 & 99.9 & 490 & 274 & 56\% & 352 & \cellcolor{heat!68}95.6 & \cellcolor{heat!6}0.0 & \cellcolor{heat!9}4.4 & 51 & \cellcolor{heat!26}31.4 & 245 & \cellcolor{heat!68}100.0 \\
\href{https://github.com/redis/redis}{\code{redis}} & de3ea25 & 374.2 & 1{,}183 & 551 & 47\% & 646 & \cellcolor{heat!6}9.1 & \cellcolor{heat!68}84.4 & \cellcolor{heat!12}6.5 & 2{,}759 & \cellcolor{heat!51}72.2 & 113 & \cellcolor{heat!6}90.3 \\
\href{https://sqlite.org/src}{\code{sqlite}} & 3.45 & 285.0 & 340 & 74 & 22\% & 131 & \cellcolor{heat!65}91.9 & \cellcolor{heat!7}1.4 & \cellcolor{heat!13}6.8 & 2{,}300 & \cellcolor{heat!49}69.4 & 150 & \cellcolor{heat!68}100.0 \\
\href{https://github.com/sudo-project/sudo}{\code{sudo}} & 1.9.15p5 & 179.8 & 354 & 79 & 22\% & 84 & \cellcolor{heat!25}35.4 & \cellcolor{heat!51}60.8 & \cellcolor{heat!8}3.8 & 187 & \cellcolor{heat!13}11.8 & 93 & \cellcolor{heat!54}97.8 \\
\href{https://github.com/tmux/tmux}{\code{tmux}} & f50a38d & 100.8 & 116 & 24 & 21\% & 302 & \cellcolor{heat!44}62.5 & \cellcolor{heat!9}4.2 & \cellcolor{heat!57}33.3 & 531 & \cellcolor{heat!19}20.2 & 27 & \cellcolor{heat!21}92.6 \\
\midrule
all &  &  &  & 2{,}078 &  &  & \cellcolor{heat!44}62.8 & \cellcolor{heat!26}27.1 & \cellcolor{heat!18}10.2 & 11{,}849 & \cellcolor{heat!44}\textbf{60.4} & 1{,}550 & \cellcolor{heat!56}\textbf{98.1} \\
\bottomrule
\end{tabular}
\end{table}

Following prior work~\cite{cheng2024sea}, we select 13 projects from OSS-Fuzz~\cite{ossfuzz} and add the widely used \code{redis}.
They range from the 12\,KLoC \code{md4c} to the 620\,KLoC \code{dovecot}.
{Following \sea, we build the ground truth (GT) by instrumenting the analyzed IR of each subject and running its test suite or a realistic workload.}
{Prior \icall\ analyses evaluate on \code{-O0} IR~\cite{lu2019where,xia2024deeptype,li2025kallgraph}.}
{We analyze each project at its default level and repeat the analysis at \code{-O0} and \code{-O2} in \S\ref{sec:eval:attribution}.}
Only observed sites are scored; \Cref{tab:subjects} reports their share as \emph{cov}, and the reference is a lower bound on the true targets.

\subsection{RQ1: Dispatch-Relation Reconstruction}\label{sec:eval:recon}

\Cref{tab:recon} measures the dispatched field $\mathit{disp}(c)$ and the assignment set $F(S{.}m)$ of \S\ref{sec:metric} directly.
Across the $2{,}078$ GT-covered sites, \ourtool\ reconstructs a dispatched-field identity at $62.8\%$ (1{,}304/2{,}078).
It classifies another $27.1\%$ (563/2{,}078) as array-indexed or global-pointer dispatches.
The remaining $10.2\%$ (211/2{,}078) stay unclassified (\S\ref{sec:boundary}).
The split varies with each subject's dispatch style: \code{redis} and \code{sudo} dispatch mostly through global function pointers, whereas \code{oniguruma}, \code{dovecot}, and \code{libjpeg-turbo} dispatch through a recovered field at over $94\%$ of sites.

No ground truth exists for $\mathit{disp}(c)$: the dynamic reference records which functions a site called and nothing about the field.
Hence, we check the recovered field against typed IR, a reference that the reconstruction never consults.
On LLVM~14 typed-pointer IR, the address computation that feeds the callee load still contains the struct type and field index that opaque IR erases.
We rebuild all 14 subjects under LLVM~14 and disassemble them with LLVM~14 tools, since a newer disassembler would upgrade the types.
We extract each \icall's dispatched field from that computation wherever it names a field of a named struct.
We match sites to the opaque-IR analysis by file, line, and column, and by inlining chain for inlined calls.
The debug information serves the typed-IR extraction only to turn a field index into a name.

Typed IR exposes such a field at $1{,}586$ sites.
The reconstruction reports a field at $1{,}550$ of them ($97.7\%$).
Of these, $1{,}520$ ($98.1\%$) agree on the field, and $1{,}516$ ($97.8\%$) agree on both struct and field.
Of the $30$ disagreements, $25$ are in \code{dovecot} and \code{redis}.
Four of these, at \code{redis}'s \code{config.c}, dispatch through a union and differ only in which union member names the location.
The one disagreement in \code{lua} is an error in the typed-IR extraction.

As \Cref{tab:recon} reports, the projection of \S\ref{sec:prune} recovers $11{,}849$ $(S.m,f)$ assignment pairs.
This assignment component has no typed-IR counterpart, so we compare it with a syntactic scan and measure its effect on pruning.
A syntactic scan that counts only stores of a function constant misses $60.4\%$ of them, which the table lists as \emph{scan misses}.
The projected set is also the better basis for pruning: under \textsc{Prune-Store} alone, its observed recall drops by $0.029$, whereas the scan's drops by $0.070$.
It also leaves fewer targets per site, $15.9$ against the scan's $19.3$.

\begin{tcolorbox}[rqbox]
\textbf{Answer to RQ1:}~\ourtool\ recovers a dispatched-field identity at $62.8\%$ of the ground-truth sites (1{,}304/2{,}078) and classifies another $27.1\%$.
Where both report the field, they agree at $98.1\%$ of sites.
Of the $11{,}849$ recovered assignment pairs, $60.4\%$ are invisible to a syntactic scan.
\end{tcolorbox}

\subsection{RQ2: Call-Graph Effectiveness}\label{sec:eval:effectiveness}

{As \S\ref{sec:gap} shows, $R_0$ fails in two ways: over-approximated sites and initially empty sites.}
We measure each repair, score the result against the ground truth, examine the edges scored as false, and report reachability and cost.

\begin{table}[t]
\begin{minipage}[t]{0.37\linewidth}
\centering
\raisebox{-\height}{\includegraphics[width=\linewidth]{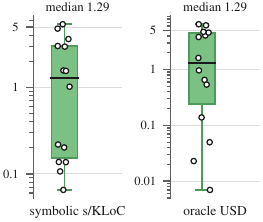}}
\Description{Two box plots over the 14 subjects, each subject a point on a log scale: symbolic analysis time in seconds per KLoC, and oracle spend in US dollars.}
\captionof{figure}{Analysis cost over the 14 subjects.}
\label{fig:cost}
\end{minipage}\hfill
\begin{minipage}[t]{0.6\linewidth}
\centering
\caption{Mean targets per site over the $N$ sites.}
\label{tab:targets}
\scriptsize\renewcommand{\arraystretch}{0.95}
\renewcommand{\code}[1]{\texttt{#1}}%
\setlength{\tabcolsep}{4pt}
\begin{tabular}{@{}lrrrrrr@{}}
\toprule
Project & $N$ & Andersen & $R_0$ & \textsc{Rules-only} & \ourtool & Ratio \\
\midrule
\code{dovecot} & 679 & \cellcolor{heat!19}50.8 & \cellcolor{heat!24}34.8 & \cellcolor{heat!41}9.1 & \cellcolor{heat!44}6.7 & \cellcolor{heat!48}5.2 \\
\code{gdbm} & 29 & \cellcolor{heat!30}20.6 & \cellcolor{heat!34}15.4 & \cellcolor{heat!46}5.9 & \cellcolor{heat!46}5.7 & \cellcolor{heat!33}2.7 \\
\code{libarchive} & 252 & \cellcolor{heat!19}50.4 & \cellcolor{heat!24}33.8 & \cellcolor{heat!42}8.3 & \cellcolor{heat!55}2.9 & \cellcolor{heat!68}11.7 \\
\code{libdwarf} & 383 & \cellcolor{heat!56}2.6 & \cellcolor{heat!56}2.6 & \cellcolor{heat!58}2.2 & \cellcolor{heat!58}2.2 & \cellcolor{heat!13}1.2 \\
\code{libjpeg-tb} & 1{,}034 & \cellcolor{heat!28}25.5 & \cellcolor{heat!30}21.5 & \cellcolor{heat!66}1.2 & \cellcolor{heat!56}2.7 & \cellcolor{heat!59}8.0 \\
\code{libxml2} & 1{,}945 & \cellcolor{heat!43}7.5 & \cellcolor{heat!47}5.6 & \cellcolor{heat!51}3.8 & \cellcolor{heat!56}2.7 & \cellcolor{heat!26}2.1 \\
\code{lua} & 19 & \cellcolor{heat!31}20.3 & \cellcolor{heat!31}19.6 & \cellcolor{heat!31}19.6 & \cellcolor{heat!38}10.9 & \cellcolor{heat!23}1.8 \\
\code{md4c} & 148 & \cellcolor{heat!64}1.4 & \cellcolor{heat!64}1.4 & \cellcolor{heat!64}1.4 & \cellcolor{heat!64}1.4 & \cellcolor{heat!9}1.0 \\
\code{nginx} & 323 & \cellcolor{heat!8}124.7 & \cellcolor{heat!13}87.8 & \cellcolor{heat!19}53.7 & \cellcolor{heat!41}8.9 & \cellcolor{heat!64}9.9 \\
\code{oniguruma} & 393 & \cellcolor{heat!36}13.7 & \cellcolor{heat!44}6.7 & \cellcolor{heat!66}1.2 & \cellcolor{heat!64}1.4 & \cellcolor{heat!46}4.8 \\
\code{redis} & 914 & \cellcolor{heat!27}28.1 & \cellcolor{heat!29}22.8 & \cellcolor{heat!38}11.1 & \cellcolor{heat!49}4.5 & \cellcolor{heat!47}5.0 \\
\code{sqlite} & 337 & \cellcolor{heat!6}148.4 & \cellcolor{heat!14}79.8 & \cellcolor{heat!29}23.4 & \cellcolor{heat!39}10.6 & \cellcolor{heat!58}7.6 \\
\code{sudo} & 278 & \cellcolor{heat!65}1.3 & \cellcolor{heat!65}1.3 & \cellcolor{heat!68}1.0 & \cellcolor{heat!64}1.4 & \cellcolor{heat!6}0.9 \\
\code{tmux} & 86 & \cellcolor{heat!24}35.2 & \cellcolor{heat!26}29.0 & \cellcolor{heat!37}12.4 & \cellcolor{heat!39}10.2 & \cellcolor{heat!33}2.8 \\
\midrule
macro & -- & \cellcolor{heat!23}37.9 & \cellcolor{heat!28}25.9 & \cellcolor{heat!38}11.0 & \cellcolor{heat!48}\textbf{5.2} & \cellcolor{heat!47}\textbf{5.0} \\
\bottomrule
\end{tabular}
\end{minipage}
\end{table}

\noindent\textbf{Over-approximated sites.}\;
\Cref{tab:targets} measures the repair at these sites by how many targets remain, which needs no ground truth.
Over the 14 subjects, the macro mean falls from $25.9$ targets per site under $R_0$ to $5.2$, a factor of $5.0$.
In comparison, \textsc{Rules-only} reduces it to $11.0$, a factor of $2.3$.
The reduction is largest where the initial sets are widest and absent where $R_0$ is already small.
\Cref{fig:size} shows the same trend in the size distribution over the ground-truth sites.
The refined call graph resolves $48.3\%$ of them to a single target, against $30.6\%$ under $R_0$.
It leaves $1.2\%$ empty, against $24.9\%$ under $R_0$.
Pruning can also empty a site that $R_0$ resolved, as at two sites on \code{nginx} and 14 on \code{sqlite}.
Each is recorded as a residual, and with a fallback to the initial set the reduction is a factor of $4.3$.

\begin{table}[t]
\centering
\caption{\ourtool\ and its ablations on the 120-site \code{nginx} ground truth. Darker cells are better within a column.}
\label{tab:ablation}
\scriptsize\renewcommand{\arraystretch}{0.95}
\setlength{\tabcolsep}{5pt}
\begin{tabular}{@{}lrrrrrrrr@{}}
\toprule
& \multicolumn{3}{c}{macro per site} & \multicolumn{2}{c}{target sets} & \multicolumn{3}{c}{edges} \\
\cmidrule(lr){2-4}\cmidrule(lr){5-6}\cmidrule(l){7-9}
Configuration & P & R & F1 & per site & empty & recovered & spurious & missed \\
\midrule
Andersen & \cellcolor{heat!6}0.037 & \cellcolor{heat!14}0.693 & \cellcolor{heat!6}0.061 & \cellcolor{heat!6}109.5 & \cellcolor{heat!8}36 & \cellcolor{heat!8}183 & \cellcolor{heat!6}12{,}960 & \cellcolor{heat!8}242 \\
Initial $R_0$ & \cellcolor{heat!8}0.049 & \cellcolor{heat!14}0.693 & \cellcolor{heat!9}0.080 & \cellcolor{heat!23}82.5 & \cellcolor{heat!8}36 & \cellcolor{heat!8}183 & \cellcolor{heat!23}9{,}717 & \cellcolor{heat!8}242 \\
\hline
\textsc{Prune-only} & \cellcolor{heat!31}0.178 & \cellcolor{heat!6}0.651 & \cellcolor{heat!34}0.244 & \cellcolor{heat!68}8.8 & \cellcolor{heat!6}37 & \cellcolor{heat!6}176 & \cellcolor{heat!68}882 & \cellcolor{heat!6}249 \\
\textsc{Candidates-only} & \cellcolor{heat!10}0.058 & \cellcolor{heat!21}0.727 & \cellcolor{heat!10}0.090 & \cellcolor{heat!20}87.5 & \cellcolor{heat!15}32 & \cellcolor{heat!11}193 & \cellcolor{heat!20}10{,}303 & \cellcolor{heat!11}232 \\
\textsc{Oracle-only} & \cellcolor{heat!13}0.075 & \cellcolor{heat!21}0.727 & \cellcolor{heat!13}0.106 & \cellcolor{heat!23}82.7 & \cellcolor{heat!15}32 & \cellcolor{heat!11}193 & \cellcolor{heat!23}9{,}736 & \cellcolor{heat!11}232 \\
\textsc{Witness-only} & \cellcolor{heat!41}0.236 & \cellcolor{heat!68}0.960 & \cellcolor{heat!39}0.282 & \cellcolor{heat!21}85.0 & \cellcolor{heat!63}4 & \cellcolor{heat!67}397 & \cellcolor{heat!22}9{,}805 & \cellcolor{heat!67}28 \\
\hline
w/o pruning rules & \cellcolor{heat!68}0.391 & \cellcolor{heat!66}0.952 & \cellcolor{heat!68}0.472 & \cellcolor{heat!66}11.6 & \cellcolor{heat!68}1 & \cellcolor{heat!68}400 & \cellcolor{heat!67}989 & \cellcolor{heat!68}25 \\
recall-preserving $\Rrp$ & \cellcolor{heat!39}0.224 & \cellcolor{heat!50}0.868 & \cellcolor{heat!35}0.255 & \cellcolor{heat!22}84.1 & \cellcolor{heat!44}15 & \cellcolor{heat!62}379 & \cellcolor{heat!23}9{,}717 & \cellcolor{heat!62}46 \\
\ourtool\ w/o LLM & \cellcolor{heat!51}0.296 & \cellcolor{heat!67}0.956 & \cellcolor{heat!50}0.355 & \cellcolor{heat!36}60.6 & \cellcolor{heat!63}4 & \cellcolor{heat!67}396 & \cellcolor{heat!37}6{,}872 & \cellcolor{heat!67}29 \\ \hline
\textbf{\ourtool} & \cellcolor{heat!66}\textbf{0.377} & \cellcolor{heat!67}\textbf{0.956} & \cellcolor{heat!66}\textbf{0.457} & \cellcolor{heat!65}\textbf{13.0} & \cellcolor{heat!68}\textbf{1} & \cellcolor{heat!68}\textbf{400} & \cellcolor{heat!67}\textbf{1{,}156} & \cellcolor{heat!68}\textbf{25} \\
\bottomrule
\end{tabular}
\end{table}

\noindent\textbf{Initially empty sites.}\;
\Cref{tab:ablation} follows the \code{nginx} graph through both repairs.
\code{nginx} has 83 initially empty sites among its 351, and 36 of them are in the ground truth.
$\Rrp$ adopts the premise-free witnesses, which give 44 of the 83 a non-empty set, whereas $\Rref$ resolves all 83.

\noindent\textbf{Precision and recall.}\;
Against the dynamic reference, the refined \code{nginx} graph reaches F1 $0.457$, up from $R_0$'s F1 of $0.080$.
Across the 14 subjects, macro F1 rises from $0.399$ to $0.708$ and recall from $0.772$ to $0.984$.
As \Cref{fig:persubject}(c,\,d) shows, the gain concentrates where type erasure breaks dispatch.
The 1{,}304 \emph{struct-field} sites gain $0.393$ precision and $0.347$ recall.
In contrast, the 547 \emph{global}-pointer sites are already precise and gain about a third as much, and the 16 array-indexed sites gain as well.
Even the 211 sites with no recovered field gain precision, at a $0.005$ recall cost.

The gain rarely harms a site: F1 improves at $1{,}150$ of the $2{,}078$ ground-truth sites, is unchanged at $911$, and worsens at $17$.
Under $\Rrp$, it improves at $189$ sites and worsens at $10$.
No subject's macro F1 falls under either call graph.
Of the 17 regressed sites, five lose no observed edge, so their F1 falls only through added targets.
The other 12 lose 14 edges, and $\Rrp$ retains all of them.
Ten of these sites lose their edges to a residual \prune, and two lose them to a \textsc{Prune-Store} premise that fails.

\noindent\textbf{Unobserved edges.}\;
Since the ground truth is a lower bound on the true targets, an unobserved edge is not necessarily wrong.
We further check every unobserved edge in two ways: against the static evidence that the analysis records and by a manual check.
\ding{182}~The refined call graph reports $10{,}354$ edges at ground-truth sites, of which $3{,}526$ are observed.
None of the other $6{,}828$ contradicts the analysis: each matches the dispatched field's signature and, where the assignment set is non-empty, is stored into that field.
{Of them, $20.3\%$ (1{,}385/6{,}828) name a function that the workload dispatched at another site.}
The rest concentrate where the assignment set merges objects: runtime-assigned fields carry $4.0$ unrefuted edges per observed edge, against $0.8$ for static tables.
\ding{183}~For the manual check, two co-authors with five years of program analysis experience independently read the source behind each of the 6{,}828 edges, 35 person-days in total.
A rater labels an edge feasible only if an assignment stores the function into the loaded field and an object that carries it can reach the site in the analyzed program.
The raters agree on $99.4\%$ of the edges (6{,}789/6{,}828), and Cohen's $\kappa$ is $0.99$.
A third co-author decides the 45 edges on which they disagree or remain uncertain.
In total, $37.4\%$ of the unobserved edges are feasible (2{,}554/6{,}828).

\noindent\textbf{Reachability and cost.}\;
At the whole-program level, the residual \add\ facts make $149$ functions reachable.
Pruning makes $951$ functions unreachable that $R_0$ reached.
\label{sec:eval:cost}\Cref{fig:cost} divides the cost between the symbolic side and the model.
The symbolic side runs once per subject, at $2.90$\,s per KLoC or $15$\,min on \code{nginx}.
The oracle cost scales with the sites queried: one analysis of the 14 subjects issues $4{,}053$ queries at \$33.9.
On \code{nginx}, the prune stage costs \ourtool\ $236$ queries and \$6.3, whereas \sea\ issues $44{,}844$ queries at \$142.8.
The difference arises because \sea\ votes on every caller--callee pair, whereas \ourtool\ issues one query per site.

\begin{tcolorbox}[rqbox]
\textbf{Answer to RQ2:}~The refined call graph cuts the mean target set from $25.9$ to $5.2$ per site without ground truth.
Against observed targets, it raises macro F1 from $0.399$ to $0.708$ and recall from $0.772$ to $0.984$.
Per site, F1 improves at $1{,}150$ of the $2{,}078$ ground-truth sites and worsens at $17$.
\end{tcolorbox}

\subsection{RQ3: Comparison with the State of the Art}\label{sec:eval:comparison}
\begin{table}[t]
\begin{minipage}[t]{0.66\linewidth}
\centering
\caption{\ourtool\ against the baselines over the 14 subjects, with the columns of \Cref{tab:ablation}.}
\label{tab:baselines}
\scriptsize\renewcommand{\arraystretch}{0.95}
\setlength{\tabcolsep}{2pt}
\begin{tabular}{@{}lllrrrrr@{}}
\toprule
& & & \multicolumn{3}{c}{macro per subject} & \multicolumn{2}{c}{target sets} \\
\cmidrule(lr){4-6}\cmidrule(l){7-8}
Resolver & Family & Input & P & R & F1 & per site & empty \\
\midrule
Initial $R_0$ & pointer & opaque IR & \cellcolor{heat!20}0.347 & \cellcolor{heat!40}0.772 & \cellcolor{heat!20}0.399 & \cellcolor{heat!14}32.2 & \cellcolor{heat!19}518 \\
TypeDive~\cite{lu2019where} & type-based & typed IR & \cellcolor{heat!30}0.402 & \cellcolor{heat!48}0.827 & \cellcolor{heat!34}0.488 & \cellcolor{heat!60}10.1 & \cellcolor{heat!59}119 \\
KallGraph~\cite{li2025kallgraph} & type-based & typed IR & \cellcolor{heat!36}0.438 & \cellcolor{heat!39}0.764 & \cellcolor{heat!39}0.520 & \cellcolor{heat!68}\textbf{6.1} & \cellcolor{heat!6}645 \\
TFA~\cite{liu2024tfa} & type-based & typed IR & \cellcolor{heat!33}0.423 & \cellcolor{heat!32}0.704 & \cellcolor{heat!34}0.492 & \cellcolor{heat!66}6.9 & \cellcolor{heat!41}290 \\
TypeCopilot~\cite{zhou2025typecopilot} & type-based & opaque IR & \cellcolor{heat!13}0.300 & \cellcolor{heat!60}0.923 & \cellcolor{heat!15}0.369 & \cellcolor{heat!6}36.3 & \cellcolor{heat!59}114 \\
\sea~\cite{cheng2024sea} & learning-based & source & \cellcolor{heat!6}0.261 & \cellcolor{heat!6}0.508 & \cellcolor{heat!6}0.312 & -- & -- \\
\ourtool\ w/o LLM & relation-based & opaque IR & \cellcolor{heat!58}0.570 & \cellcolor{heat!63}0.945 & \cellcolor{heat!58}0.642 & \cellcolor{heat!50}14.7 & \cellcolor{heat!58}124 \\
\hline
\textbf{\ourtool} & relation-based & opaque IR & \cellcolor{heat!68}\textbf{0.631} & \cellcolor{heat!68}\textbf{0.984} & \cellcolor{heat!68}\textbf{0.708} & \cellcolor{heat!65}7.7 & \cellcolor{heat!68}\textbf{24} \\
\bottomrule
\end{tabular}
\end{minipage}\hfill
\begin{minipage}[t]{0.32\linewidth}
\centering
\raisebox{-\height}{\includegraphics[width=\linewidth]{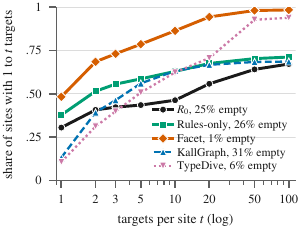}}
\Description{Cumulative curves over ground-truth sites: the share of sites resolved to between one and t targets for the initial call graph, the rules alone, Facet, KallGraph, TypeDive, and TypeCopilot, with each resolver's share of empty sites in the legend.}
\captionof{figure}{Ground-truth sites resolved to at most $t$ targets.}
\label{fig:size}
\end{minipage}
\end{table}

\begin{figure}[t]
\centering
\includegraphics[width=0.58\linewidth]{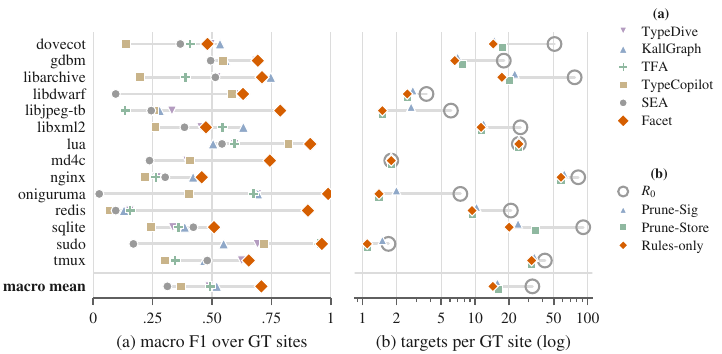}\hfill
\includegraphics[width=0.4\linewidth]{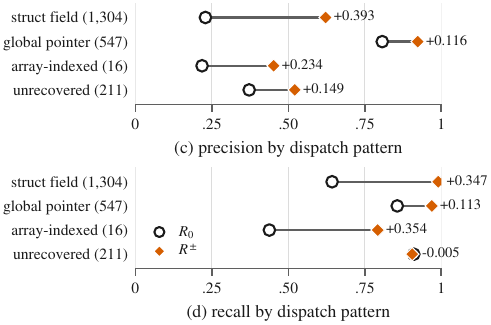}
\Description{Two dot plots sharing one column of subject names: on the left, macro F1 per subject for TypeDive, KallGraph, TFA, TypeCopilot, SEA and Facet; on the right, targets per ground-truth site under $R_0$, each pruning rule alone, and the two composed, on a log axis. Right of them, two short dumbbell plots give precision and recall by dispatch pattern (struct field, global pointer, array-indexed, unrecovered) under R_0 and the refined call graph.}
\caption{Per-subject results: macro F1 per resolver (a), targets per ground-truth site under $R_0$, each prune rule, and both (b), and precision and recall by dispatch pattern from $R_0$ to $\Rref$ (c, d).}
\label{fig:persubject}
\end{figure}

\Cref{tab:baselines} and \Cref{fig:persubject}(a) compare \ourtool\ with every baseline over the 14 subjects.
\ourtool\ achieves the highest macro F1, $0.708$, against $0.520$ for the best baseline, KallGraph.
The comparison has three parts: type analysis on typed IR, recovered types on opaque IR, and an LLM-based analysis.
We then compare \ourtool\ with KallGraph edge by edge.

To rule out that \ourtool's margin comes only from the type-based tools running without their types, we further examine two settings in which those tools have the types.
\emph{Type analysis on typed IR} gives these tools the types directly.
On the LLVM~14 typed-pointer rebuilds, \mlta\ reaches macro F1 $0.488$, KallGraph $0.520$, and TFA $0.492$.
All three fall below \ourtool's $0.708$ on opaque IR.
On \code{nginx}, typed-IR \mlta\ still misses $220$ targets that \ourtool\ recovers, and an initializer witnesses $192$ of them.
Hence, what \mlta\ lacks is the evidence of which functions the program assigns to the field.
{The second setting, \emph{recovered types on opaque IR}, tests whether re-inferring the types recovers even that much.}
TypeCopilot~\cite{zhou2025typecopilot}'s multi-layer mode scores macro F1 $0.369$, below the $0.488$ of \mlta\ on typed IR.
Its released configuration scores $0.052$, because matching on type sets leaves most sites empty.
Hence, re-inferred types do not reach what typed IR gives \mlta.
Even without the model, \ourtool\ exceeds every typed-IR tool (\S\ref{sec:eval:attribution}).

The \emph{LLM-based analysis} \sea~\cite{cheng2024sea} reaches macro F1 $0.312$ under the Gemini model that we use and \sea's documented five-vote protocol.
\sea\ reports targets by definition site, so \Cref{tab:baselines} leaves its per-site columns blank.
Most of the gap arises before the model is queried, because \sea\ only filters the candidates of its parser.
The parser admits $48.8\%$ of the ground-truth sites and loses the dispatch hidden behind macro expansion.
Even on the admitted sites, \sea\ reaches macro F1 $0.509$.

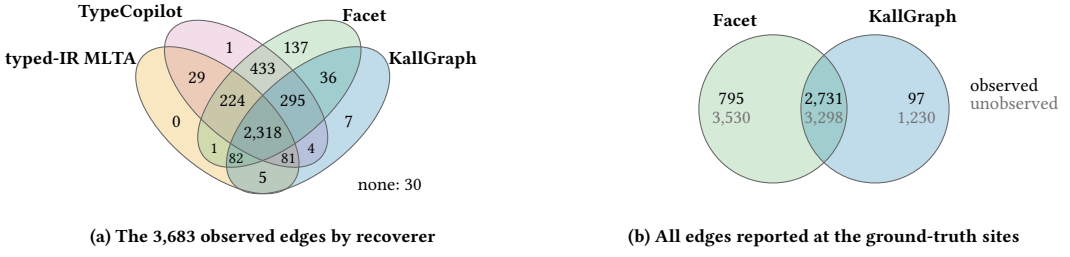
\begin{figure}[t]
\centering
\providecommand{\mlta}{MLTA}
\definecolor{vFacet}{RGB}{124,193,132}\definecolor{vKG}{RGB}{0,114,178}\definecolor{vMLTA}{RGB}{230,159,0}\definecolor{vTC}{RGB}{204,121,167}
\begin{tikzpicture}[x=1mm,y=1mm,font=\scriptsize,
  circ/.style={draw=black!60,line width=0.4pt},
  n/.style={inner sep=0pt,align=center},
  sl/.style={font=\scriptsize\bfseries,inner sep=0pt},
  capt/.style={font=\scriptsize\bfseries,inner sep=0pt,anchor=north,align=center}]
\begin{scope}[shift={(0,0)}]
\fill[vMLTA,fill opacity=0.25,rotate around={-40:(14.5,8.5)}] (14.5,8.5) ellipse (13 and 6.5);
\fill[vTC,fill opacity=0.25,rotate around={-40:(18.3,12.0)}] (18.3,12.0) ellipse (13 and 6.5);
\fill[vFacet,fill opacity=0.32,rotate around={40:(22.7,12.0)}] (22.7,12.0) ellipse (13 and 6.5);
\fill[vKG,fill opacity=0.25,rotate around={40:(26.5,8.5)}] (26.5,8.5) ellipse (13 and 6.5);
\draw[circ,rotate around={-40:(14.5,8.5)}] (14.5,8.5) ellipse (13 and 6.5);
\draw[circ,rotate around={-40:(18.3,12.0)}] (18.3,12.0) ellipse (13 and 6.5);
\draw[circ,rotate around={40:(22.7,12.0)}] (22.7,12.0) ellipse (13 and 6.5);
\draw[circ,rotate around={40:(26.5,8.5)}] (26.5,8.5) ellipse (13 and 6.5);
\node[n] at (9.1,8.4) {0};
\node[n] at (16.1,18.2) {1};
\node[n] at (24.9,18.2) {137};
\node[n] at (31.9,8.4) {7};
\node[n] at (11.8,14.2) {29};
\node[n] at (29.2,14.2) {36};
\node[n] at (20.5,15.3) {433};
\node[n] at (16.5,11.1) {224};
\node[n] at (24.5,11.1) {295};
\node[n] at (20.5,6.7) {2{,}318};
\node[n] at (20.5,0.9) {5};
\node[n,font=\tiny] at (14.1,4.8) {1};
\node[n,font=\tiny] at (26.9,4.8) {4};
\node[n,font=\tiny] at (17.0,3.5) {82};
\node[n,font=\tiny] at (23.9,3.5) {81};
\node[sl,anchor=east] at (3.9,16.5) {typed-IR \mlta};
\node[sl,anchor=south east] at (10.0,21.5) {TypeCopilot};
\node[sl,anchor=south west] at (31.0,21.5) {Facet};
\node[sl,anchor=west] at (37.1,16.5) {KallGraph};
\node[n,anchor=west] at (33.0,0.0) {none: 30};
\node[capt] at (20.5,-5.8) {(a) The 3{,}683 observed edges by recoverer};
\end{scope}

\begin{scope}[shift={(72,0)}]
\coordinate (F) at (16,10); \coordinate (K) at (29.5,10);
\fill[vFacet,fill opacity=0.32] (F) circle (9.8); \fill[vKG,fill opacity=0.25] (K) circle (9.8);
\draw[circ] (F) circle (9.8); \draw[circ] (K) circle (9.8);
\node[sl,anchor=south] at (11,20.8) {Facet};
\node[sl,anchor=south] at (34.5,20.8) {KallGraph};
\node[n] at (10.5,10) {795\\[-1pt]\textcolor{black!55}{3{,}530}};
\node[n] at (22.75,10) {2{,}731\\[-1pt]\textcolor{black!55}{3{,}298}};
\node[n] at (35,10) {97\\[-1pt]\textcolor{black!55}{1{,}230}};
\node[n,anchor=west,align=left] at (42,12) {observed\\[-1pt]\textcolor{black!55}{unobserved}};
\node[capt] at (22.75,-5.8) {(b) All edges reported at the ground-truth sites};
\end{scope}
\end{tikzpicture}%
\Description{Two Venn diagrams. (a) The 3,683 observed edges split by which of four resolvers recover them (Facet, KallGraph, typed-IR MLTA, TypeCopilot): all four 2,318; Facet and TypeCopilot only 433; Facet, TypeCopilot and KallGraph 295; Facet, TypeCopilot and MLTA 224; Facet alone 137; Facet, KallGraph and MLTA 82; KallGraph, MLTA and TypeCopilot 81; Facet and KallGraph 36; MLTA and TypeCopilot 29; KallGraph alone 7; KallGraph and MLTA 5; KallGraph and TypeCopilot 4; TypeCopilot alone 1; Facet and MLTA 1; MLTA alone 0; none 30. (b) All edges reported at the ground-truth sites by Facet and KallGraph: both report 2,731 observed and 3,298 unobserved edges; Facet alone 795 observed and 3,530 unobserved; KallGraph alone 97 observed and 1,230 unobserved.}
\caption{Edge-set overlaps between \ourtool\ and the baselines.}
\label{fig:venn}
\end{figure}
Finally, as \Cref{fig:venn}a shows, all four resolvers recover $2{,}318$ of the $3{,}683$ observed edges.
\ourtool\ alone recovers $137$ edges, against $7$ for KallGraph alone, $1$ for TypeCopilot alone, and none for \mlta\ alone.
TFA recovers no edge that the other four miss.
Another $433$ edges are recovered only by \ourtool\ and TypeCopilot.
However, TypeCopilot reports $36.3$ targets per site, against $7.7$ for \ourtool.
Of the $433$, $397$ lie at the $398$ expansion sites of \code{redis}'s \code{REDISMODULE\_GET\_API} macro.
Without these sites, \ourtool's macro F1 is $0.690$ and still exceeds KallGraph's $0.543$.
Of the $157$ observed edges \ourtool\ misses, $139$ are absent from $R_0$ and $5$ are removed by a warrant.

KallGraph keeps objects apart by tracing each function to the call, but leaves a site empty wherever the trace breaks: $645$ sites, against $24$ for \ourtool.
\ourtool\ instead merges all objects of one type.
Accordingly, KallGraph reports fewer targets at $261$ sites.
Of these, $207$ dispatch through runtime-assigned fields whose objects KallGraph separates and \ourtool\ merges.
As \Cref{fig:venn}b shows, this merging reduces precision: each of the $795$ observed edges that only \ourtool\ recovers comes with $4.4$ unobserved edges, and each of KallGraph's $97$ with $13$.

\smallskip

\begin{tcolorbox}[rqbox]
\textbf{Answer to RQ3:}~\ourtool\ outperforms every baseline of the three families: its macro F1 is $0.708$, against $0.520$ for the best baseline, KallGraph.
\ourtool\ and KallGraph are complementary: their union resolves $2{,}049$ of the $2{,}078$ sites.
\end{tcolorbox}

\subsection{RQ4: Ablation Study}\label{sec:eval:attribution}

As \Cref{tab:baselines} shows, the symbolic and neural parts of \ourtool\ fail differently on their own.
The symbolic part alone reaches macro F1 $0.642$ but leaves $124$ ground-truth sites empty and $14.7$ targets per site, because a warrant can act only where the program supplies evidence.
An LLM alone, as \sea\ uses it, reaches $0.312$ under the same model.
Together they reach $0.708$, and the $0.066$ that the model adds is $21\%$ of the gain over $R_0$.

As \Cref{fig:persubject}(b) shows, \emph{the prune warrants} narrow the target set at a small recall cost.
\textsc{Prune-Sig} halves it, from $32.2$ to $15.7$ targets per ground-truth site, without losing recall on any subject.
\textsc{Prune-Store} narrows it by a similar amount, to $15.9$, and the two rules together reach $14.3$.
\textsc{Prune-Store} loses four observed edges on three subjects, at most $0.025$ recall on \code{gdbm}.
Each lost edge comes from a store that the projection cannot attribute to a field: a literal aggregate, a call result, or an array-element temporary.
Admitting such stores by signature alone would recover all four at $0.1$ more targets per site, but \ourtool\ does not admit them.
On \code{nginx}, the two rules also lower F1 from $0.472$ without them to $0.457$ (\Cref{tab:ablation}).
\emph{The add warrants} supply most of the recovered recall.
On \code{nginx}, the witnesses alone raise the recovered observed edges from $183$ to $397$, close to the $400$ of the full analysis.
Over the 14 subjects, they add $797$ observed edges, against $129$ from the oracle.
\textsc{Witness-only} reaches recall $0.960$, above the refined call graph's $0.956$, because it never prunes.

\emph{The oracle} increases the recovered observed edges from 3{,}410 to 3{,}526.
It reduces the unobserved edges from 21{,}101 to 6{,}828 and the empty ground-truth sites from 124 to 24.
It improves per-site F1 at 214 sites and worsens it at 11.
Gains are largest without field evidence (\code{sudo}: F1 0.557 to 0.962) and negligible where witnesses suffice.
For additions, the oracle is queried only where witnesses leave a site empty and is the sole source at 108 sites across the 14 subjects.
On \code{nginx}, it resolves the four remaining sites from 7.2 candidates each, versus 149 from signature matching.
A populated but incomplete site is never queried, which leaves 45 of the $3{,}683$ observed edges missing.
Two checks, precedence and abstention, constrain the changes.
Disabling precedence lowers the refined graph's recall below that of $\Rrp$ on four subjects.
Abstention blocks changes at 609 sites with one or two symbolic targets, withholding 949 witness additions and 205 model removals.
Among the 336 ground-truth sites in this group, only 9 of the 430 withheld additions were observed, whereas 20 of the 23 blocked removals would have deleted observed edges.
{\Cref{fig:sensitivity}(a) separates the effect of abstention from the choice of its threshold.}
{Setting $\tau$ to 0 turns abstention off and lowers macro F1 by $0.031$.}
{Replaying the provenance with $\tau$ from 1 to 4 changes macro F1 by at most $0.003$.}

\noindent\textbf{Comparison of the Two Call Graphs.}
\Cref{tab:ablation} evaluates both call graphs on the same classified facts on \code{nginx}.
$\Rrp$ adopts only the 326 initializer-witnessed \add\ facts, and $\Rref$ also accepts the premised warrants and the residual facts.
Dropping the residual \add\ facts from the provenance gives precision $0.361$, recall $0.939$, and F1 $0.441$.
{The premised warrants thus account for the precision gain over $\Rrp$'s $0.224$, and the residual additions for the remaining recall up to $\Rref$'s $0.956$.}
Across the 14 subjects, $\Rrp$ adds $2{,}745$ edges and removes none; $452$ of them are ground-truth edges that $R_0$ missed.
At ground-truth sites, $37.6\%$ of the initializer-witnessed additions are observed (452/1{,}203), against $15.9\%$ of the store-witnessed (345/2{,}165).

\begin{table}[t]
\begin{minipage}[t]{0.52\linewidth}
\centering
\raisebox{-\height}{\includegraphics[width=\linewidth]{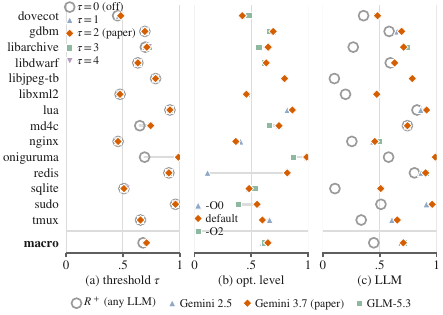}}
\Description{Three range plots sharing one column of subject names plus a macro-mean row. Each row is a horizontal bar from the smallest to the largest value of one factor, with a marker per setting and a red diamond for the paper's setting: (a) macro F1 of the refined call graph at tau from 0 to 4, with tau=0 as a hollow circle; (b) macro F1 of Facet without the LLM at -O0, the default level, and -O2; (c) macro F1 of the refined call graph under Gemini 2.5, Gemini 3.7, and GLM-5.3, with the recall-preserving graph as a hollow circle.}
\captionof{figure}{Sensitivity to the threshold $\tau$ (a), the optimization level (b), and the LLM (c).}
\label{fig:sensitivity}
\end{minipage}\hfill
\begin{minipage}[t]{0.47\linewidth}
\centering
\caption{The 17 deep bugs reported upstream.}
\label{tab:bugs}
\scriptsize
\renewcommand{\arraystretch}{0.85}\setlength{\aboverulesep}{0.30ex}\setlength{\belowrulesep}{0.45ex}%
\setlength{\tabcolsep}{2pt}
\renewcommand{\code}[1]{\texttt{#1}}
\begin{tabular}{@{}lllr@{}}
\toprule
Project & Bug type (CWE) & Report & Status \\
\midrule
\multirow{5}{*}{\href{https://github.com/dovecot/core}{\code{dovecot}}} & double free (415) & private & reported \\
 & incorrect comparison (697) & private & confirmed \\
 & use after free (416) & private & confirmed \\
 & incorrect calculation (682) & PR\,319 & reported \\
 & premature EOF (393) & PR\,320 & reported \\
\midrule
\multirow{2}{*}{\href{https://github.com/libarchive/libarchive}{\code{libarchive}}} & infinite loop (835) & \#3492 & confirmed \\
 & uninitialized read (457) & \#3496 & confirmed \\
\midrule
\href{https://github.com/davea42/libdwarf-code}{\code{libdwarf}} & \code{sizeof} on a pointer (467) & PR\,325 & confirmed \\
\midrule
\multirow{2}{*}{\href{https://github.com/libjpeg-turbo/libjpeg-turbo}{\code{libjpeg-tb}}} & out-of-bounds write (787) & \#915 & confirmed \\
 & out-of-bounds read (125) & \#916 & confirmed \\
\midrule
\href{https://gitlab.gnome.org/GNOME/libxml2}{\code{libxml2}} & type confusion (843) & MR\,463 & confirmed \\
\midrule
\multirow{3}{*}{\href{https://git.kernel.org/pub/scm/linux/kernel/git/torvalds/linux.git}{\code{linux}}} & out-of-bounds read (125) & private & confirmed \\
 & out-of-bounds read (125) & private & reported \\
 & out-of-bounds write (787) & private & reported \\
\midrule
\multirow{2}{*}{\href{https://github.com/nginx/nginx}{\code{nginx}}} & resource exhaustion (400) & \#1732 & confirmed \\
 & interpretation conflict (436) & H1\,4014106 & confirmed \\
\midrule
\href{https://github.com/tmux/tmux}{\code{tmux}} & NULL dereference (476) & private & confirmed \\
\bottomrule
\end{tabular}
\end{minipage}
\end{table}

\noindent\textbf{Optimization Sensitivity.}\;
{\Cref{fig:sensitivity}(b) tests whether the optimization level that the projects ship with favors \ourtool: we rebuild every subject at \code{-O0} and at \code{-O2}, trace the ground truth on each rebuild, and repeat the analysis without the LLM.}
The gain over $R_0$ stays between $0.247$ and $0.275$ at every level.
{Individual subjects change by up to $0.171$, except \code{redis} at \code{-O0}, whose F1 falls from $0.816$ to $0.115$.}
{Its 398 \code{REDISMODULE\_GET\_API} sites dispatch through a pointer that the source assigns through an integer cast.}
{At \code{-O0}, the cast remains a \code{ptrtoint} and \code{inttoptr} pair, which Andersen's analysis does not propagate through, so these sites are empty in $R_0$ and stay empty.}
Without debug information, no fact attaches to a site: stripping the debug metadata of \code{gdbm} and \code{lua} makes the refined call graph equal $R_0$.

\noindent\textbf{Model Sensitivity.}
{As \Cref{fig:sensitivity}(c) shows, we repeat the analysis under three models:} Gemini~2.5~Flash~\cite{gemini2025report}, Gemini~3.7~Flash, and GLM-5.3~\cite{glm53card} from a second vendor.
$\Rrp$ is identical under all three, because it admits no LLM decision.
The refined call graph's macro F1 is $0.689$, $0.708$, and $0.714$ under the three models.
This span of $0.025$ compares with a gap of $0.309$ between the refined call graph and $R_0$ (\Cref{tab:baselines}).
The widest per-subject span is $0.059$, on \code{nginx}.
Equal scores do not imply equal decisions: on \code{oniguruma}, the three models propose $1{,}683$, $1{,}994$, and $2{,}071$ \prune\ facts but score identically, because a residual \prune\ acts only on edges that no warrant covers (\S\ref{sec:adopt}).

\begin{tcolorbox}[rqbox]
\textbf{Answer to RQ4:}~The symbolic part alone reaches macro F1 $0.642$, above every prior tool, but leaves $124$ ground-truth sites empty.
The LLM resolves $100$ of these sites, halves the target set, and removes $68\%$ of the remaining spurious edges, which raises F1 to $0.708$.
$\Rrp$ is identical under three LLMs from two vendors.
\end{tcolorbox}

\subsection{RQ5: Application to Bug Detection}\label{sec:eval:bugs}

\Cref{tab:bugs} lists the 17 bugs that the refined call graph helped find across the 14 subjects and the Linux kernel.
The audit starts from the sites that $R_0$ leaves empty and the refined call graph resolves.
$R_0$ leaves $24.9\%$ of the ground-truth sites empty (518/2{,}078).
At each audited site, we derive from the caller the contract that every target must satisfy, and we inspect the implementation that disagrees with its siblings.
We reported each bug upstream with a fix, and seven of these reports went through private responsible disclosure.
Maintainers have confirmed 12 of the 17 so far.

\code{libjpeg-turbo} illustrates the audit in \Cref{lst:jpeg}.
The decompressor dispatches upsampling through the \code{\_upsample} field of \code{jpeg\_upsampler} at line~2.
The object comes from the library's memory manager and the field is set by runtime stores at lines~5--7, so $R_0$ and $\Rrp$ resolve nothing here.
In contrast, the refined call graph resolves exactly the three stored implementations under the runtime-store warrant.
Their caller, \code{jpeg\_skip\_scanlines()}, determines the contract: after a skip, the next delivered row is the first row after it.
The implementations disagree at the early return of line~11 when a skip does not start on a row-group boundary.
Then either a row from before the skip is delivered, or \code{rows\_to\_go} is unchanged and decoding runs past \code{output\_height} into the caller's buffer.
Across 71{,}040 decodes, 883 outputs were wrong and nine hung (details in issue \#915).

\Cref{lst:ext4} shows one of the three kernel bugs.
All three are deep: they lie on \code{ext4}'s \code{EXT4\_IOC\_GROUP\_ADD} resize path.
They sit eight frames below the \code{f\_op->unlocked\_ioctl} \icall\ at line~2, which is the only entry from user space to that path.
$R_0$ leaves that site empty and the refined call graph resolves it to \code{ext4\_ioctl} among six others.
The bug shown tests its bound at line~12, after the read.
Hence, the loop reads one entry past the block, and the count it returns can equal the block's capacity.
The caller uses that count as an index into the same block at line~17: the resize succeeds while writing out of bounds.
The Kernel Address Sanitizer (KASAN) confirms all three on Linux~6.12.9 and~7.2.4.
Two have been present since \code{ext4} was split from \code{ext3} in 2006, and the third since Linux~3.3.

\begin{figure}[]
\begin{minipage}[]{0.49\linewidth}
\begin{lstlisting}[basicstyle=\ttfamily\fontsize{6}{6.8}\selectfont,xleftmargin=0pt,aboveskip=2pt,belowskip=2pt,breaklines=true,breakindent=0pt,morekeywords={_upsample,sep_upsample,merged_2v_upsample,merged_1v_upsample,rows_to_go,output_scanline,output_height,lines_left_in_iMCU_row,num_lines,increment_simple_rowgroup_ctr,jpeg_skip_scanlines,master,using_merged_upsample},label=lst:jpeg,caption={The \code{libjpeg-turbo} case.}]
/* jdpostct.c:156 */
(*cinfo->upsample->_upsample)(cinfo, input_buf,
                       in_row_group_ctr, ...);
/* jdsample.c:467, jdmerge.c:485, jdmerge.c:531 */
upsample->pub._upsample = sep_upsample;
upsample->pub._upsample = merged_2v_upsample;
upsample->pub._upsample = merged_1v_upsample;
/* jdapistd.c:581, jpeg_skip_scanlines() */
if (num_lines < lines_left_in_iMCU_row) {
  increment_simple_rowgroup_ctr(cinfo, num_lines);
  return num_lines;   // group not drained; rows_to_go stays stale
} else {
  cinfo->output_scanline += lines_left_in_iMCU_row;
  if (!master->using_merged_upsample)
    upsample->rows_to_go =
      cinfo->output_height - cinfo->output_scanline;
}
\end{lstlisting}
\end{minipage}\hfill%
\begin{minipage}[]{0.49\linewidth}
\begin{lstlisting}[basicstyle=\ttfamily\fontsize{6}{6.8}\selectfont,xleftmargin=0pt,aboveskip=2pt,belowskip=2pt,breaklines=true,breakindent=0pt,morekeywords={unlocked_ioctl,ext4_ioctl,verify_reserved_gdb,reserve_backup_gdb,ext4_list_backups,gdbackups,EXT4_ADDR_PER_BLOCK,EXT4_BLOCKS_PER_GROUP,le32_to_cpu,cpu_to_le32},label=lst:ext4,caption={The Linux \code{ext4} case.}]
/* fs/ioctl.c:51 */
error = filp->f_op->unlocked_ioctl(filp, cmd, arg);
/* fs/ext4/file.c:979 */
.unlocked_ioctl = ext4_ioctl,
/* fs/ext4/resize.c:788, verify_reserved_gdb(): the
   block holds EXT4_ADDR_PER_BLOCK(sb) entries */
while ((grp = ext4_list_backups(sb, ...)) < end) {
  if (le32_to_cpu(*p++) !=
      grp * EXT4_BLOCKS_PER_GROUP(sb) + blk) {
    return -EINVAL;   // p[APB] read before ...
  }
  if (++gdbackups > EXT4_ADDR_PER_BLOCK(sb))
    return -EFBIG;    // ... the counter stops
}
return gdbackups;       // largest returned: APB
/* fs/ext4/resize.c:1094, reserve_backup_gdb() */
data[gdbackups] =
  cpu_to_le32(blk + primary[i]->b_blocknr);
\end{lstlisting}
\end{minipage}
\Description{Two code listings: the libjpeg-turbo dispatch through the _upsample field, its three stored implementations, and the skip-scanlines caller; and the ext4 ioctl dispatch with the verify_reserved_gdb loop that reads one entry past the block.}
\end{figure}

\smallskip
\begin{tcolorbox}[rqbox]
\textbf{Answer to RQ5:}~The audit starts from the sites that the refined call graph resolves and the initial call graph $R_0$ leaves empty.
It reported 17 bugs upstream, each with a fix, and 12 are confirmed.
The three bugs in Linux \code{ext4} had been latent since 2006 or for at least 14 years.
\end{tcolorbox}

\section{Discussion}\label{sec:discussion}
\subsection{Limitations}\label{sec:boundary}

\ding{182} The reconstruction takes receiver types from debug information.
Hence, it reaches only receivers that a debug record names or a typed field chain leads to (\S\ref{sec:disp}).
At the remaining $10.2\%$ of the ground-truth sites, only the model changes the initial set.
Without debug information, the refined call graph equals $R_0$ (\S\ref{sec:eval:attribution}).
Link-time optimization is untested.
The assignment relation is field-based rather than field-sensitive~\cite{moller2026spa,pearce2007field}, which leaves $\Rref$ wider where objects of one type hold different targets (\S\ref{sec:eval:comparison}).
\ding{183} Completion acts only at initially empty sites, so a populated but incomplete set is never queried.
Hence, 45 of the $3{,}683$ observed edges remain missing (\S\ref{sec:eval:attribution}).

\subsection{Threats to Validity}\label{sec:threats}

\noindent\textbf{External threats.}\;\ding{182}~\emph{Generalizability.}
Subject selection may affect the generalizability of our results, because the 14 programs may not represent C practice.
To mitigate this, we selected the subjects following prior work~\cite{cheng2024sea}, and they cover the dispatch patterns of \Cref{tab:recon} in varying proportions.
Additionally, we extended the audit of \S\ref{sec:eval:bugs} to the Linux kernel.
\ding{183}~\emph{Data leakage.}
We mitigate it in two ways.
{First, the LLM never sees the ground truth, and the recall-preserving call graph $\Rrp$ admits no model decision.}
Second, we replaced every field name in the prompt with an opaque token, which changes macro F1 over the sites that the oracle resolved from $0.792$ to $0.790$.

\noindent\textbf{Internal threats.}\;\ding{182}~\emph{Ground truth.}
As in prior work~\cite{cheng2024sea,liu2024tfa,li2025kallgraph}, we build the reference dynamically, so it is a lower bound on the true targets.
To mitigate this, we report two measurements that do not depend on the reference: the target reduction (\S\ref{sec:eval:effectiveness}) and the agreement of the reconstruction with typed IR (\S\ref{sec:eval:recon}), and two co-authors reviewed every edge that the reference counts as false (\S\ref{sec:eval:effectiveness}).
\ding{183}~\emph{Premises of the prune rules.}
{\textsc{Prune-Store} and \textsc{Prune-Sig}} rest on premises that C code can violate.
To mitigate this, $\Rrp$ uses neither rule, and where $\Rref$ accepts them, the recall loss is at most $0.025$ (\S\ref{sec:eval:attribution}).
\ding{184}~\emph{Non-determinism of the LLM.}
To mitigate this, the oracle runs at temperature~0, and three models from two vendors change the macro F1 of $\Rref$ by at most $0.025$ (\S\ref{sec:eval:attribution}).
\ding{185}~\emph{Threshold.}
The threshold $\tau$ could have been tuned to the subjects.
To mitigate this, $\tau$ is \ourtool's only threshold and was fixed before the evaluation; replaying it from 1 to 4 changes macro F1 by at most $0.003$ (\S\ref{sec:eval:attribution}).

\section{Related Work}\label{sec:related}
\subsection{Indirect-Call Analysis}

Prior analyses key a site to its targets in three ways: by type, by a type that they recover once the IR no longer contains one, or by what a model decides.
\Cref{tab:baselines} scores all three families against \ourtool.

\noindent\textbf{Type-based resolution.} These analyses key a site to the functions of a compatible type, and they differ in which type the IR still contains.
\flta~\cite{tice2014forward,ge2016fine} keys on the signature alone, and \mlta~\cite{lu2019where} on the declared type of the loaded field, which DeepType~\cite{xia2024deeptype}, TyPM's module dependence~\cite{lu2023typm}, and Kelp's regional pointer information~\cite{cai2024kelp} refine.
{TFA~\cite{liu2024tfa} rebuilds the erased struct type from debug metadata and byte offsets, which our recovery also uses, and co-analyzes the rebuilt type with value flow.}
What separates \ourtool\ from them is which fact it rebuilds: they rebuild the type that keys the match, but \ourtool\ rebuilds the assignment relation that the type encoded.
KallGraph~\cite{li2025kallgraph} is the nearest in aim: it replaces \mlta's type-level relation with an object-level one, traced from each store to the call site on typed IR.
\ourtool\ joins on the field instead, which needs neither typed IR nor a complete path but merges the objects of a type (\S\ref{sec:eval:comparison}).

\noindent\textbf{Resolution on opaque IR.} TypeCopilot~\cite{zhou2025typecopilot} is the only prior \icall\ analysis for opaque IR.
It re-infers the erased pointee types so that \mlta\ can match on them again.

\noindent\textbf{Learned resolution.} \sea~\cite{cheng2024sea} asks an LLM whether a caller summary matches a callee summary and keeps the majority answer over repeated queries.
{Zhu et al.~\cite{zhu2023callee} learn the same decision for binaries, where neither debug layouts nor a points-to solution is available.}
Both adopt the model's answer as an edge, with at most a majority vote over repeated queries, because no program evidence checks it (\S\ref{sec:gap}).
Instead, \ourtool\ classifies the model's facts by the rules that classify the symbolic ones, and the warrants take precedence over the model's facts.

\subsection{Other Related Work}

\ourtool\ refines the call graph of an inclusion-based points-to analysis~\cite{andersen1994program,smaragdakis2015pointer,hardekopf2007ant} in its value-flow implementation~\cite{sui2016svf}.
A stronger solver~\cite{yao2026phoenix} would improve the initial graph that the warrants refine.
However, a structure-sensitive solver keys objects on the typed pointers that opaque IR removes~\cite{balatsouras2016structure}.
Our prune warrants use the complement argument of call-graph refinement over IFDS~\cite{milanova2004precise,reps1995ifds}.
In Java, the counterpart of indirect calls is reflection.
Analyses resolve reflective calls from strings, casts, or runtime logs~\cite{livshits2005reflection,li2019tosem,landman2017reflection}.
In contrast, a C function pointer carries none of this information.
The soundiness manifesto~\cite{livshits2015soundiness} and defensive points-to analysis~\cite{smaragdakis2018defensive} ask what an analysis that omits dynamic features such as reflection can still guarantee.
Our two call graphs answer this question.
A probabilistic abstract interpretation instead quantifies precision~\cite{shi2026confidence}.
Other work combines learning with static analysis: neural outputs have been combined with logical reasoning~\cite{li2023scallop}, and LLMs can provide information that static analyzers lack~\cite{kan2026specguru}.
Classifiers remove edges from Java call graphs without explaining each removal~\cite{utture2022striking,lecong2022autopruner}.

\section{Conclusion}\label{sec:conclusion}

We presented \ourtool, an indirect-call analysis for LLVM IR with opaque pointers.
Rather than recovering the erased pointee types, it reconstructs the dispatch relation from debug layouts and program writes.
Every refinement is classified under an explicit warrant, which leaves only the residual cases to an LLM.
On 14 C programs, it reaches macro F1 $0.708$ against the best prior tool's $0.520$, and its refined call graph led an audit to 17 deep bugs.
Two lessons extend beyond this setting: a relation that a type once encoded can be reconstructed without the type, and a refinement that records its evidence and premise lets each client decide how much to trust.
Future work will feed admitted facts back into the points-to solution and cover C++ virtual dispatch.

\bibliographystyle{ACM-Reference-Format}
\bibliography{ref}
\end{document}